\documentclass{aa}  

\usepackage{graphicx}
\usepackage{txfonts}
\usepackage{hyperref}
\begin{document} 

\title{Observational evidence for a possible link between PAH emission and dust trap locations in protoplanetary disks}

\author{Nienke van der Marel\inst{1}
  \and Niels F. W. Ligterink \inst{2,3}
  \and Ryan van der Werf \inst{1}
\and Milou Temmink \inst{1}
\and Paola Pinilla \inst{4}
  \and Bin Jia \inst{1}
  \and Quincy Bosschaart \inst{1}
}

\institute{Leiden Observatory, Leiden University, Leiden, The Netherlands,
\and Faculty of Aerospace Engineering, Delft University of Technology, Delft, The Netherlands, 
\and Center for Space and Habitability, University of Bern, Gesellschaftsstrasse 6, CH-3012 Bern, Switzerland
\and Mullard Space Science Laboratory, University College London, Dorking, UK}

\authorrunning{van der Marel et al.}
\titlerunning{A link between PAH emission and dust traps in protoplanetary disks}
\date{Accepted for publication in A\&A}

\abstract{Polycyclic Aromatic Hydrocarbons (PAHs) are commonly detected in protoplanetary disks, but it is unclear what causes the wide range of intensities across the samples.} {In this work, the measured PAH intensities  of a range of disks are compared with ALMA dust continuum images, in order to test whether there is evidence that PAHs are frozen out on pebbles in dust traps and only sublimate under certain conditions.} {A sample is constructed from 26 T Tauri and Herbig disks located within 300 pc, with constraints on the 3.3 $\mu$m PAH intensity  and with high-resolution ALMA continuum data. The midplane temperature is derived using a power-law or with radiative transfer modeling. The warm dust mass is computed by integrating the flux within the 30 K radius and convert to a dust mass.}{A strong correlation with a Pearson coefficient of 0.88$\pm$0.07 between the 3.3 $\mu$m PAH intensity and the warm dust mass was found. The correlation is driven by the combination of deep upper limits and strong detections corresponding to a range of warm dust masses. Possible correlations with other disk properties like FUV radiation field or total dust mass are much weaker. Correlations with PAH features at 6.2, 8.6 and 11.3 $\mu$m are potentially weaker, but this could be explained by the smaller sample for which these data were available.} {The correlation is consistent with the hypothesis that PAHs are generally frozen out on pebbles in disks, and are only revealed in the gas phase if those pebbles have drifted towards warm dust traps inside the 30 K radius and vertically transported upwards to the disk atmosphere  with sufficiently high temperature to sublimate PAHs into the gas phase. This is similar to previous findings on complex organic molecules in protoplanetary disks and provides further evidence that the chemical composition of the disk is governed by pebble transport.}


\maketitle

\section{Introduction}
Polycyclic Aromatic Hydrocarbons (PAHs) are molecules that are constituted of two or more integrated aromatic rings, often set in a honeycomb structure of carbon atoms, locking up $\sim20$\% of all elemental carbon \citep[e.g.][]{Visser2007} and emit several spectral features through vibrational transitions at infrared wavelengths \citep{Allamandola1985}, at e.g. 3.3, 6.2, 7.7, 8.6, 11.2 and 12.7 $\mu$m. PAHs are strong absorbers in the ultraviolet (UV) and an important contributor to the local gas heating. The different features are attributed to various C-H and C-C bonds as well as ionization \citep[e.g.][]{Acke2010}. PAHs are generally found in high UV environments and have been observed in a wide variety of sources \citep[the interstellar medium, dense molecular clouds, circumstellar envelopes and (proto-)planetary nebulae, ][]{Peeters2004}. However, the appearance of PAHs in protoplanetary disks, where the building blocks of comets and planets are formed, is less well understood, in particular the wide variation of PAH spectral intensities in disks. 
Dozens of protoplanetary disks have been targeted with infrared spectroscopy with e.g. ISO, \emph{Spitzer} and VLT \citep{Habart2004,Geers2006,Geers2007,Acke2010}, and very recently, with JWST \citep[e.g.][]{Sturm2024}. These surveys generally show infrared PAH signatures in around 60-70\% of the spectra of Herbig stars and only 8\% for T Tauri stars. Although a general correlation between PAH detection and stellar temperature linked to the UV radiation field exists, it is not clear why the PAH emission strength or even its detection also varies strongly within spectral type and millimeter-dust disk morphology as traced by ALMA. Specifically, some Herbig disks show relatively strong PAH emission when their disk has a cavity (e.g. HD100546), whereas others do not (e.g. HD135344B), while multi-ringed disks without cavity have much weaker PAH emission (e.g. HD163296), but   compact disks show strong PAH emission (e.g. HD179218). For T Tauri stars there is a similar range of PAH emission strength across morphologies \citep{Geers2007}. These findings hint at a possible link between the millimeter-dust substructures and the PAH intensity.

Several mechanisms have been proposed to explain the diversity of PAH emission in protoplanetary disks: flared Herbig disks (known as Group I disks) have a higher PAH emission due to their larger exposure to UV radiation compared to the flat, Group II disks due to the absorption of PAHs by the inner dusty regions in Group II disks \citep{Meeus2001,Habart2004}; the lack of PAH emission in many T Tauri stars may be related to the destruction of PAHs due to EUV and X-ray photons \citep{Siebenmorgen2010} or efficient freeze out onto dust grains \citep{Geers2009}, followed by ionisation reactions which deplete the PAHs \citep{Bouwman2010}; PAHs cluster in outer disk regions where the UV radiation is lower and therefore PAH dissociation is limited, so that PAH emission at short wavelengths gets suppressed \citep{Rapaciola2009,Lange2021}. 

The latter led to a more thorough investigation combining cluster formation with freeze-out and vertical mixing \citep[Figure 1 in][]{Lange2023}, showing that clustering followed by freeze-out in the low UV regime near the midplane of the disk can indeed deplete gas-phase PAHs by a factor 50-1000 compared to the ISM. This suggests a scenario where PAHs are locked up in the icy layers on dust grains, unless the dust grains are warm enough to release their icy cover.

\begin{figure}[!t]
    \centering
    \includegraphics[width=0.5\textwidth]{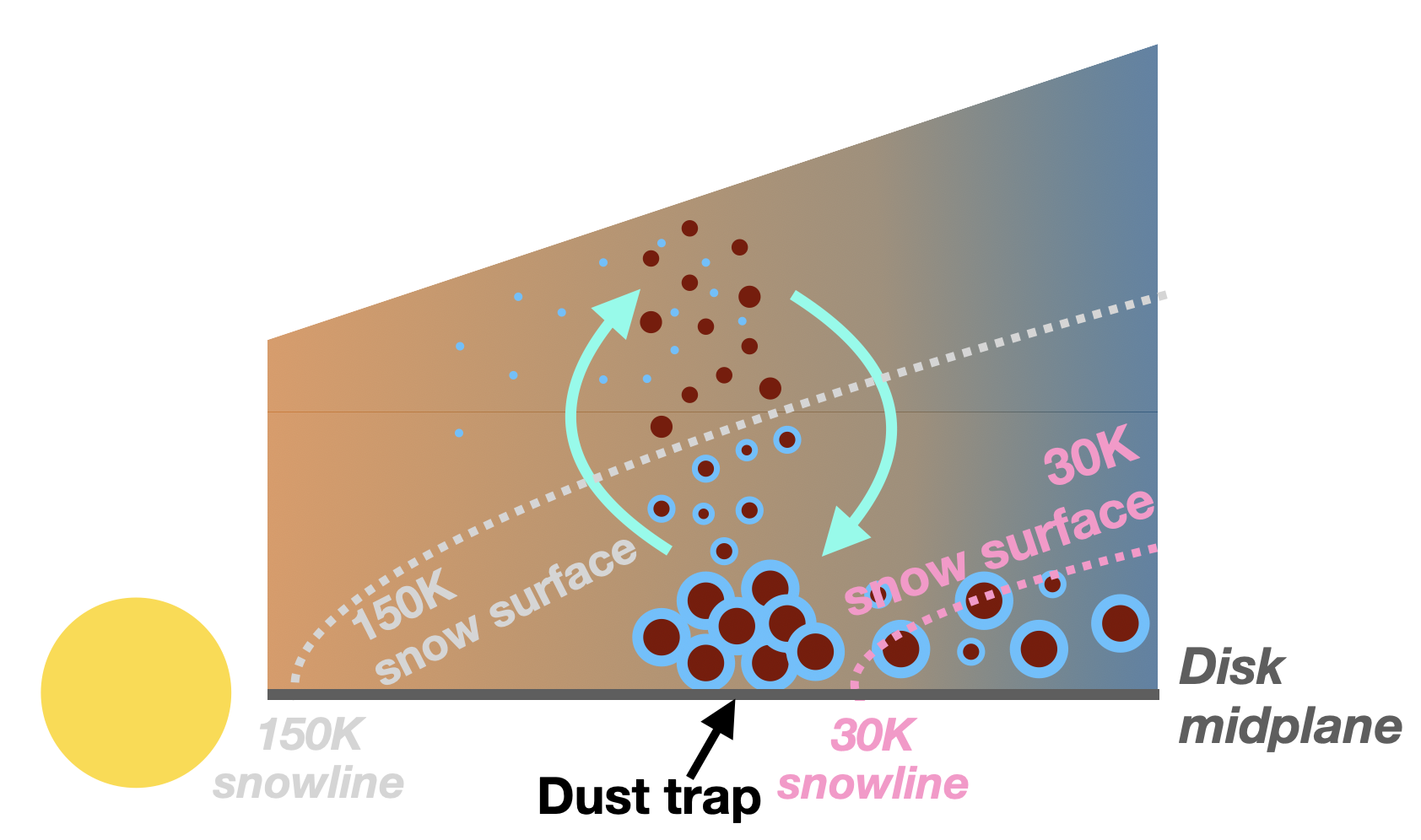}
    \caption{Graphic of the proposed scenario of sublimation in a disk with a warm dust trap, viewed from the side, based on \citet{vanderMarel2021-irs48}. If the bulk of the dust content is located in a dust trap inside the CO snowline at 30 K, this implies that fragments of its icy pebbles can be transported vertically to the higher disk layers above the H$_2$O snow surface at 150 K, and sublimate their icy content into the gas-phase, where it becomes observable. If the bulk of the pebbles is located outside the CO snowline, its icy layers may remain frozen out and its contents are not revealed in the gas phase. Such a scenario may also apply to PAHs, which is the hypothesis of this work.}
    \label{fig:graphic}
\end{figure}

In recent years, ALMA observations have demonstrated that a similar mechanism is at play for complex organic molecules (COMs) in disks, in particular in so-called dust traps. Dust traps are recognized as substructures observed in millimeter observations of protoplanetary disks \citep[e.g.][]{Andrews2020} and are generally interpreted as the result of pressure bumps, which halt millimeter-sized dust grains from drifting inwards \citep{Pinilla2012b}. In IRS~48, HD~100546 and HD100453 gas-phase COMs with sublimation temperatures $>$100 K have been found to be highly abundant and cospatial with these dust traps \citep[e.g.][]{Booth2021-hd100,vanderMarel2021-irs48,Brunken2022,Booth2024,Booth2025}. This is interpreted as the release of COMs into the gas-phase due to vertical mixing of the small grains after fragmentation of millimeter-sized pebbles at the location of the dust trap, where the COMs were locked up in icy mantles of the grains and pebbles \cite[see Figure 3 in][]{vanderMarel2021-irs48}. Vertical mixing is required, as the midplane temperature is generally too low at the dust trap location for sublimation, whereas the higher molecular layers in the disk are significantly warmer \citep{Bruderer2012}. The icy pebbles have been transported inwards toward the dust trap through radial drift and are inherited from the original molecular cloud. This means that the classical picture of a radial snowline is revised to a `snow surface', i.e. a snowline that is varying with both temperature and height in the disk \citep[e.g][]{Gavino2021}, where the radius of the midplane CO snowline lies below the H$_2$O (and CH$_3$OH) snow surface (see Figure \ref{fig:graphic}). It has been proposed that other disks do not show any gas-phase COM emission as their dust traps are located too far out in the disk, i.e. their dust traps are considered `cold', whereas IRS~48 and HD~100546 happen to have `warm' dust traps \citep{vanderMarel2023,Temmink2023}. A similar scenario has been proposed to explain the dichotomy in C/O ratios in disks as measured from C$_2$H emission \citep{vanderMarel2021-c2h}. The chemical complexity of irradiated icy dust traps may even lead to the efficient formation of IOM (insoluble organic matter) or macromolecules,  a dominant reservoir of organic carbon in meteorites \citep{Alexander2008,Ligterink2024}.

Spatially resolved PAH emission of a handful of disks using ground-based facilities with $\sim$0.1-0.2" point spread function (PSF) indicated initially that PAH features were centrally peaked, while long wavelength features were found to be more extended than those at shorter wavelengths \citep{vanBoekel2004,Habart2004-hd97048,Habart2006,Lagage2006,Geers2007-irs48,Bouteraon2019}. However, at least for PAH features at longer wavelengths the emission has now often been shown to be ring-like \citep{Devinat2022,Yoffe2023}. These spatially resolved features are not obviously linked to the dust substructures, but the interpretation remains limited due to the small sample size. Also, these surveys have only targeted disks where PAHs have been previously detected, so non-detections of PAHs in other disks with similar properties remain unexplained. Therefore, in this study we consider the integrated PAH intensities and upper limits on those for a much larger disk sample, and compare their values with the amount of warm dust in these disks.

Within this study, `warm dust' is defined as the millimeter dust inside the CO snowline, i.e. where CO ice has been sublimated and the midplane temperature is $>$30 K. Although PAHs sublimate at much higher temperatures $>$150-300 K \citep{Ligterink2023}, such temperatures are easily reached at a few scale heights above the midplane \citep{vanderMarel2021-irs48},
so efficient vertical transport of fragmented icy grains to warmer layers where ices could sublimate is required.  
The definition of the cutoff temperature of the warm dust region is thus chosen at 30 K in the midplane. 

In this study we will investigate the correlation between warm dust and PAH gas-phase intensity using a sample of suitable protoplanetary disks for which PAH intensities are available from the literature, as well as high-resolution ALMA observations and well-defined stellar properties. The paper is structured as follows: Section \ref{sec:data} presents the sample selection criteria and the collection of ALMA data, as well as details on the ALMA observations and temperature profiles. Section \ref{sec:analysis} explains the methods that were used to compute the warm dust mass. Section \ref{sec:results} presents the resulting correlations. Section \ref{sec:discussion} discusses the implications of our results and draws the conclusions.

\begin{table*}[!ht]
    \centering
    \caption{Sample of disks in this study, based on PAH studies by \citet{Habart2004} and \citet{Geers2006}.}
    \label{tbl:sample}
    \begin{tabular}{l|llllll|ccl|l}
    \hline
         Target & $d$ & $L_*$ & SpT & $M_*$ & $\log\dot{M}$ & $L_{FUV}^a$ & $F_{PAH3.3\mu{m}}$ &  $I_{PAH3.3\mu{m},1pc}$ & Disk type$^b$ & Ref$^c$\\
         &(pc)&($L_{\odot}$)&&($M_{\odot}$)&($M_{\odot}$ yr$^{-1}$)&($L_{\odot}$)&(10$^{-14}$ W m$^{-2}$)&&&\\
         \hline
ABAur & 155 & 46&A0&2.4& -7.0 & 3.5 & $<$1.0$\pm$0.2 & $<$3.0$\pm$0.6 & TD & 2  \\ 
Elias1/V892Tau	&	134	&	129$^d$	&	A0	&	2	&	-	&	7.8	&	0.5$\pm$0.1	&	0.87$\pm$0.18	&	TD	&	1	\\
HD100453	&	104	&	6.2	&	A9	&	1.5	&	-8.3	&	0.1	&	1.3$\pm$0.2	&	1.8$\pm$0.3	&	TD	&	3	\\
HD100546	&	110	&	23	&	B9	&	2.2	&	-7	&	2.5	&	2.5$\pm$0.5	&	3.8$\pm$0.8	&	TD	&	2	\\
HD135344	&	136	&	6.2	&	F5	&	1.6	&	-7.4	&	0.1	&	$<$0.5$\pm$0.1	&	$<$1.2$\pm$0.2	&	TD	&	2	\\
HD139614	&	135	&	5.9	&	A9	&	1.5	&	-8.1	&	0.1	&	$<$0.5$\pm$0.1	&	$<$1.1$\pm$0.2	&	TD	&	3	\\
HD142527	&	157	&	9.1	&	F6	&	2.3	&	-7.5	&	0.1	&	1.0$\pm$0.6	&	3.1$\pm$1.9	&	TD	&	2	\\
HD169142	&	114	&	20	&	A5	&	2	&	-8.7	&	0.56	&	1.0$\pm$0.2	&	1.6$\pm$0.3	&	TD	&	2	\\
HD97048	&	185	&	35	&	A0	&	2.4	&	-8.2	&	2.1	&	1.3$\pm$0.3	&	5.6$\pm$1.3	&	TD	&	2	\\
IRS48	&	134	&	18	&	A0	&	2	&	-8.4	&	1.1	&	0.24$\pm$0.02	&	0.54$\pm$0.04	&	TD	&	2	\\ 
SR21	&	138	&	12	&	G3	&	2	&	-7.9	&	0.04	&	0.35$\pm$0.02	&	0.84$\pm$0.05	&	TD	&	2	\\ 
SYCha	&	181	&	0.73	&	M0.5	&	0.8	&	-9.4	&	0.0004	&	$<$0.008$\pm$0.002	&	$<$0.033$\pm$0.006	&	TD	&	2	\\
TCha	&	102	&	1.3	&	G8	&	1.2	&	-8.4	&	0.01	&	$<$0.25$\pm$0.05	&	$<$0.33$\pm$0.06	&	TD	&	2	\\
GWLup	&	155	&	0.33	&	M1.5	&	0.5	&	-9.0	&	0.0008	&	$<$0.024$\pm$0.005	&	$<$0.072$\pm$0.015	&	RD	&	5	\\
HD142666	&	148	&	8.7	&	A8	&	1.6	&	-7.6	&	0.2	&	0.3$\pm$0.2	&	0.83$\pm$0.55	&	RD	&	3	\\
HD163296	&	101	&	17	&	A1	&	2.0	&	-7.4	&	1	&	$<$0.2$\pm$0.04	&	$<$0.26$\pm$0.05	&	RD	&	5	\\
IMLup	&	158	&	2.6	&	K5	&	0.9	&	-7.9	&	0.009	&	$<$0.19$\pm$0.04	&	$<$0.60$\pm$0.12	&	RD	&	5	\\
RULup	&	159	&	1.5	&	K7	&	0.6	&	-7.1	&	0.04	&	$<$0.57$\pm$0.11	&	$<$1.8$\pm$0.4	&	RD	&	5	\\
V1121Oph/AS209	&	121	&	1.4	&	K5	&	0.8	&	-7.3	&	0.04	&	0.12$\pm$0.02	&	0.22$\pm$0.04	&	RD	&	5	\\ 
WaOph6	&	123	&	2.9	&	K6	&	0.7	&	-6.6	&	0.12	&	0.23$\pm$0.02	&	0.44$\pm$0.04	&	RD	&	5	\\
DoAr24E	&	136	&	1.6	&	K0	&	0.6	&	-8.5	&	0.004	&	$<$0.041$\pm$0.008	&	$<$0.095$\pm$0.019	&	CD	&	4	\\
GQLup	&	150	&	0.91	&	K5	&	0.6	&	-7.4	&	0.03	&	$<$0.054$\pm$0.011	&	$<$0.15$\pm$0.03	&	CD	&	4	\\
Haro1-17	&	142	&	0.22	&	M7	&	0.1	&	-99	&	0	&	$<$0.0045$\pm$0.001	&	$<$0.011$\pm$0.002	&	CD	&	4	\\
HD104237/DXCha	&	108	&	21	&	A4	&	1.8	&	-6.4	&	1.5	&	$<$0.5$\pm$0.1	&	$<$0.73$\pm$0.14	&	CD	&	3	\\
HD179218	&	266	&	112	&	A0	&	3	&	-	&	6.8	&	1.7$\pm$0.2	&	15$\pm$1.8	&	CD	&	6	\\
Sz73	&	155	&	0.18	&	K7	&	0.7	&	-8.5	&	0.006	&	0.094$\pm$0.02	&	0.28$\pm$0.06	&	CD	&	4	\\
WXCha	&	189	&	0.86	&	M0	&	0.5	&	-6.7	&	0.1	&	$<$0.027$\pm$0.006	&	$<$0.12$\pm$0.03	&	CD	&	4	\\
         \hline
    \end{tabular}\\
         $^a$ FUV luminosity, integrated from the stellar spectrum (including accretion luminosity) between 912-2050 $\AA$ or 6-13.6 eV \citep{Habing1968}. In Figure \ref{fig:PAH_other} $L_{\rm FUV}$ is computed at 150 au radius to get the parameter $\chi$, similar to previous studies.\\
         $^b$ TD = transition disk, CD = compact disk ($<$20 au), RD = ring disk (see \citet{vdMMulders2021}).\\
         $^c$ Reference for stellar properties: 1) \citet{Herczeg2014} 2) \citet{vanderMarel2023}, 3) \citet{Grant2023}, 4) \citet{Manara2023}, 5) \citet{Andrews2018}, 6) \citet{Vioque2018}. \\
         $^d$ Whereas \citet{Vioque2018} reports a luminosity of 1.3 $L_{\odot}$, the higher luminosity by \citet{Herczeg2014} is used here, as it is more consistent with the stellar mass, which is independently derived from CO kinematics by \citet{Long2021}.
\end{table*}

\section{Data}
\label{sec:data}
\subsection{Sample}
The sample consists of a selection of targets from \citet{Habart2004} and \citet{Geers2007}, who present ISO spectra of Herbig stars and VLT spectra of T Tauri stars, respectively. These studies report (amongst others) the integrated intensity of the 3.3 $\mu$m PAH feature. 
From their original target lists, the targets at distances $>$300 pc are excluded as their dust disk cannot be well constrained due to a lack of spatially well-resolved ALMA data. Targets in the high extinction CrA region are excluded due to lack of proper stellar information. We also remove close binaries with separation $\sim$0.5-1" (EM SR 9, HT Lup and SX Cha) as dust disks in such binary systems are truncated and less representative of pure radial drift effects \citep[e.g.][]{Akeson2014}. Furthermore, we exclude  
WL16 as it is an embedded disks and HD141569 as it is an evolved, debris-like disk. This leaves 26 targets, which are listed in Table \ref{tbl:sample}.

\subsection{Stellar properties and PAH intensities}
Stellar properties are taken from the recent literature, which often includes updated values based on the Gaia DR2 distances compared to the original papers.

PAH emission consists of multiple infrared features, at e.g. 3.3, 6.2, 7.7, 8.6, 11.2 and 12.7 $\mu$m. The integrated 3.3 $\mu$m PAH feature is most commonly reported across a wide range of spectral types, both detections and non-detections (upper limits), and is thus chosen for our main sample analysis. The origin of this feature is likely due to a single aromatic material and its intensity ratio with aliphatic features is independent of the local UV-field, suggesting that these PAHs are continuously replenished at the disk surface \citep{Bouteraon2019}. The measured PAH flux as well as the distance-scaled PAH intensity are listed in Table \ref{tbl:sample}. The latter is computed by multiplying the flux with 4$\pi d^2\times10^{-5}$ with $d$ the distance in pc. The (scaled) uncertainties are included as well. The upper limits are 5$\sigma$, following the convention of the original works. 

In addition, we consider the 6.2, 8.6 and 11.3 $\mu$m PAH intensities derived by \citet{Habart2004} and \citet{Acke2010}, as these are bright, commonly studied features, and the 8.6 and 11.3 features have been followed up with spatially resolved observations \citep[e.g.][]{Yoffe2023}. Whereas the 3.3 $\mu$m feature is dominated by emission from the inner region of the disk, PAH features at longer wavelengths are known to be more extended as they get a larger contribution from the outer disk, due to the lower UV radiation needed to excite these features \citep{Habart2004}. Furthermore, the 11.3 $\mu$m feature is thought to be related to neutral PAHs, whereas the 8.6 $\mu$m feature is linked to ionized PAHs, and these features are thus not tightly correlated with each other \citep[e.g][]{Maaskant2014}. Only 12 of our targets (all Herbigs) overlap with the sample from \citet{Acke2010}, which means that the analysis of these features is much more limited. This subsample and their scaled PAH intensities are listed in Table \ref{tab:pahacke}, where the intensities are scaled to the stellar distance, as explained above.

\begin{table}[h]
    \caption{Subsample of targets with long-wavelength PAH intensities, original data from \citet{Acke2010}}
    \label{tab:pahacke}
    \centering
    \begin{tabular}{l|lll}
    \hline
    Target & $I_{PAH6.2\mu{m},1pc}$&$I_{PAH8.6\mu{m},1pc}$&$I_{PAH11.3\mu{m},1pc}$ \\
    \hline
         AB Aur&  14.3 & 7.2 & 3.3\\
         HD100453 & 2.9 & 0.64 & 1.3 \\
         HD100546 & 22$^a$ & 3.5 & $<$9.1 \\
         HD135344 & 0.86 & 0.13 & 0.6 \\
         HD139614 & 1.6 & 0.57 & $<$0.11 \\
         HD142527 & 6.7 & 1.6 & 5.0 \\
         HD169142 & 4.2 & 1.5 & 3.4 \\
         HD97048 & 23.7 & 12.8 & 27.2 \\
         HD142666 & 2.0 & 0.7 & 1.2 \\
         HD163296 & $<$1.9$^a$ & $<$1.2 & $<$2.6 \\
         HD104237 & $<$0.06 & $<$0.9 & $<$1.3 \\
         HD179218 & 72$^a$ & 22 & $<$27 \\
         \hline
    \end{tabular}\\
    $^a$ 6.2 $\mu$m flux taken from \citet{Habart2004}.\\
\end{table}

\begin{table*}[!ht]
    \centering
    \caption{Properties ALMA continuum data used in this study}
    \label{tbl:almadata}    
    \begin{tabular}{llllllllllll}
    \hline
         Target & ALMA dataset & Frequency & Beam size & Total flux & $M_{\rm dust,total,20K}$ & $M_{\rm warm dust,>30K}$& $R_{in}^a$&$R_{size}^b$&Origin$^a$  \\
         &&(GHz)&(")&(mJy)&($M_{\oplus}$)&($M_{\oplus}$)&(au)&(au)&\\
    \hline   
ABAur & 2012.1.00303.S & 338 & 0.31x0.17 & 181 & 45$\pm$5 & 26$\pm$3 & 150 & 214 & 0 \\
Elias1/V892Tau	&	2013.1.00498.S	&	225	&	0.23x0.16	&	284	&	120$\pm$12	&	73$\pm$7	&	29	&	54	&	1	\\
HD100453	&	2017.1.01424.S	&	281	&	0.03x0.02	&	218	&	39$\pm$4	&	23$\pm$2	&	30	&	42	&	2	\\
HD100546	&	2016.1.00344.S	&	225	&	0.04x0.02	&	397	&	145$\pm$15	&	87$\pm$9	&	25	&	72	&	2	\\
HD135344B	&	2016.1.00340.S	&	155	&	0.09x0.06	&	43	&	67$\pm$7	&	33$\pm$3	&	52	&	91	&	3	\\
HD139614	&	2022.1.01302.S	&	226	&	0.16x0.12	&	164	&	90$\pm$9	&	34$\pm$3	&	19	&	70	&	4	\\
HD142527	&	2012.1.00631.S	&	322	&	0.12x0.09	&	898	&	659$\pm$66	&	243$\pm$24	&	185	&	238	&	2	\\
HD169142	&	2016.1.00344.S	&	225	&	0.05x0.03	&	135	&	53$\pm$5	&	32$\pm$3	&	24	&	67	&	2	\\
HD97048	&	2016.1.00826.S	&	338	&	0.06x0.03	&	2061	&	730$\pm$73	&	167$\pm$17	&	63	&	226	&	2	\\
IRS48	&	2013.1.00100.S	&	343	&	0.18x0.13	&	163	&	29$\pm$3	&	17$\pm$2	&	70	&	98	&	5	\\
SR21	&	2018.1.00689.S	&	217	&	0.05x0.05	&	96	&	55$\pm$6	&	33$\pm$3	&	56	&	66	&	6	\\
SYCha	&	2018.1.00689.S	&	217	&	0.12x0.08	&	58	&	57$\pm$6	&	3$\pm$0.3	&	35	&	298	&	7	\\
TCha	&	2015.1.00979.S	&	97	&	0.08x0.04	&	15	&	51$\pm$5	&	8$\pm$1	&	34	&	54	&	2	\\
GWLup	&	DSHARP	&	231	&	0.04x0.03	&	83	&	51$\pm$5	&	3$\pm$0.3	&	0.04	&	97	&	9	\\
HD142666	&	DSHARP	&	232	&	0.03x0.02	&	131	&	80$\pm$8	&	43$\pm$4	&	0.21	&	54	&	9	\\
HD163296	&	DSHARP	&	239	&	0.05x0.04	&	634	&	166$\pm$17	&	64$\pm$6	&	0.29	&	130	&	9	\\
IMLup	&	DSHARP	&	231	&	0.04x0.04	&	209	&	134$\pm$13	&	20$\pm$2	&	0.11	&	242	&	9	\\
RULup	&	DSHARP	&	231	&	0.03x0.02	&	195	&	127$\pm$13	&	25$\pm$3	&	0.09	&	79	&	9	\\
V1121Oph/AS209	&	DSHARP	&	231	&	0.04x0.04	&	233	&	88$\pm$9	&	8$\pm$1	&	0.08	&	127	&	9	\\
WaOph6	&	DSHARP	&	231	&	0.06x0.05	&	156	&	61$\pm$6	&	15$\pm$2	&	0.12	&	91	&	9	\\
DoAr24E	&	2016.1.00336.S	&	225	&	0.11x0.09	&	6	&	4$\pm$0.4	&	2$\pm$0.2	&	0.09	&	6	&	10	\\
GQLup	&	2015.1.00773.S	&	240	&	0.05x0.02	&	24	&	14$\pm$1.4	&	5$\pm$0.5	&	0.07	&	30	&	8	\\
Haro1-17	&	2013.1.00157.S	&	344	&	0.16x0.14	&	9	&	2$\pm$0.2	&	0.1$\pm$0.01	&	0.03	&	9	&	11	\\ 
HD104237/DXCha	&	2021.1.01137.S	&	225	&	0.17x0.10	&	88	&	31$\pm$3	&	18$\pm$2	&	0.32	&	9	&	12	\\
HD179218	&	2021.1.00709.S	&	105	&	0.24x0.19	&	5	&	87$\pm$9	&	55$\pm$6	&	0.74	&	21	&	12	\\
Sz73	&	2018.1.01458.S	&	226	&	0.08x0.06	&	11	&	8$\pm$1	&	$<$8	&	0.03	&	5	&	12	\\
WXCha	&	2021.1.00854.S	&	233	&	0.45x0.22	&	9	&	10$\pm$1	&	$<$10	&	0.06	&	$<$19	&	12	\\

    \hline     
    \end{tabular}\\
    $^a$ Inner edge dust disk: either cavity radius for transition disks or dust sublimation radius (where T$>$1500 K) for compact/ring disks \\
    $^b$ Dust disk size, see text for definition.
    $^c$ References published ALMA images: 0) \citet{vanderMarel2021-asymm}, 1) \citet{Pinilla2018}, 2) \citet{Francis2020}, 3) \citet{Cazzoletti2018}, 4) Bosschaart et al. subm., 5) \citet{vanderMarel2021-irs48}, 6) \citet{Yang2023}, 7) \citet{Orihara2023}, 8) \citet{Wu2017}, 9) \citet{Andrews2018}, 10) \citet{Ansdell2020}, 11) \citet{Cox2017}, 12) Product fits file.
\end{table*}

\subsection{ALMA data}
\label{sec:ALMA}
High-resolution ALMA dust continuum images are available for all targets (for the image gallery, see Figure \ref{fig:gallery}). The majority of these images have been previously published, and we use the published images for this work. For 5 targets we use the product fits files directly from the ALMA archive as these are sufficient for our purposes. All information on the ALMA images are listed in Table \ref{tbl:almadata}. The disks are classified as either a transition disk (TD) with an inner dust cavity, a ring disk (RD) with multiple dust rings (but no inner cavity) or a compact disk (CD) with no evidence for dust rings and a dust disk radius $<$20 au, following the classification by \citet{vdMMulders2021}. The ALMA continuum images are used to derive the radial intensity profile of the dust continuum, integrated flux and dust disk size, as explained in the next section.

The comparison of the ALMA disk morphology with the scaled PAH intensities immediately reveals that the two are not uniquely linked, but that the spectral type plays a role as well: 5 of the 13 transition disks have an upper limit of the 3.3 PAH feature, both for early and late type stars; the 7 ring disks have both upper limits and (weak) detections, irrespective of spectral type and for the compact disks only 1 disk has a strong PAH detection, whereas this is not the only early type star. Therefore, the combination of disk morphology and spectral type likely plays a role.

\begin{figure*}

    \includegraphics[width=\textwidth]{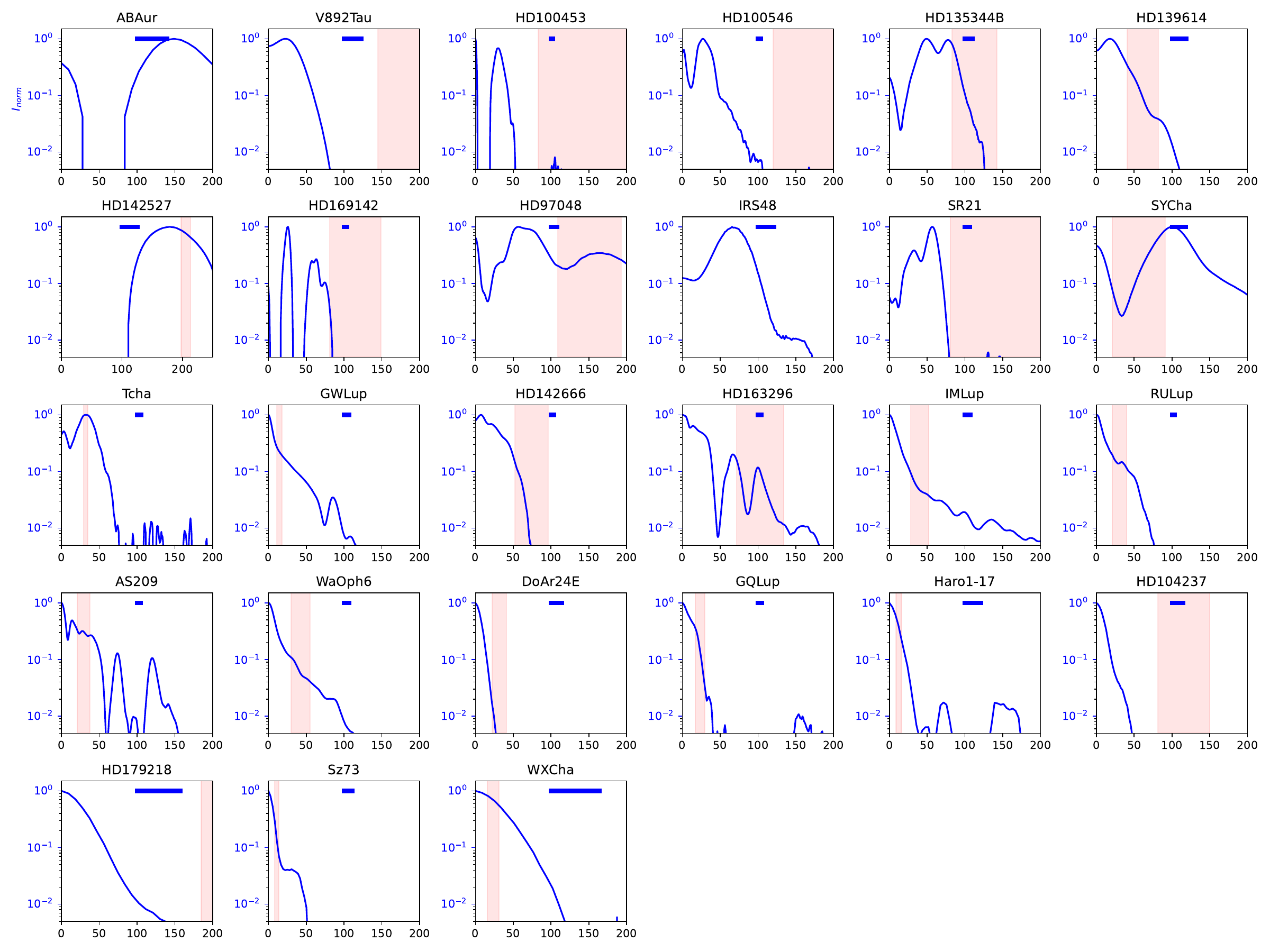}
\caption{Azimuthally averaged profiles of the ALMA continuum images (see Figure \ref{fig:gallery}) of the targets in this study. The profiles are normalized to the peak and shown in logarithmic scale. The beam size is indicated with a horizontal blue bar in the top right of each profile. The pink shaded region shows the CO snowline in each disk, as defined by the 22-30 K temperature regime, based on the derived temperature profile (see text). For IRS48, the snowline is beyond the shown radial range. The part of the profile to the left of the pink region is considered the 'warm dust region' where at least part of the ice is sublimated, potentially releasing PAHs.}
\label{fig:profiles}
\end{figure*}

\section{Analysis}
\label{sec:analysis}
In the first step of the analysis, we check whether the PAH intensity at 3.3 $\mu$m is correlated with the stellar luminosity, FUV radiation field or disk dust mass. The FUV radiation field is computed from the stellar spectrum and the accretion rate, using the Kurucz models in combination with a 10,000 K blackbody scaled to the accretion luminosity, following the method by \citet{Kama2016}. The FUV luminosity $L_{FUV}$ is computed by integrating the stellar spectrum between 912-2050 $\AA$. The FUV radiation field across the disk, named $\chi$, is estimated from $L_{FUV}$ at a distance of 150 au from the star, scaled to the interstellar radiation field ISRF (1.6$\cdot10^{-6}$ W m$^{-2}$ \citep{Habing1968}, following the approach from previous works \citep{Habart2004}, where a correlation with this parameter is equivalent to a correlation with the FUV luminosity itself. Furthermore, we estimate the FUV radiation field at the inner edge of the disk $R_{in}$, which is either the cavity radius for the transition disks or the dust sublimation radius (defined as $R_{\rm sub}\approx0.07(L_*/L_{\odot})^{1/2}$, assuming $T_{\rm sub}=$1500 K, \citet{Dullemond2001}) for other disks, as well as the radiation field at the outer edge of the dust disk, taken as the size of the dust disk $R_{\rm size}$. 

The dust disk size $R_{\rm size}$ is derived from the ALMA images (Figure \ref{fig:gallery}) using a curve-of-growth method, encircling 90\% of the total flux, following previous works \citep{Tripathi2017}. The compact disks are only marginally resolved, so for those disks we use the CASA \texttt{imfit} task to estimate the FWHM to measure the radius of each dust disk. For the majority of targets $R_{\rm size}$ is measured at $\sim$1 mm wavelength (ALMA Band 6 or 7), only for HD135344B, T Cha and HD179218 longer wavelength images were used, which means that their disk sizes could be somewhat larger at 1 mm. These radii are listed in Table \ref{tbl:almadata}.

Finally, we estimate the total dust mass using the common equation related to the total millimeter flux $F_{\nu}$, for an assumed temperature of 20 K and optically thin emission \citep{Hildebrand1983}:
\begin{equation}
\label{eqn:mdust}
M_{dust}=\frac{F_{\nu}d^2}{\kappa_{\nu}B_{\nu}(T)}    
\end{equation}
with distance $d$, Planck function $B_{\nu}$, dust opacity $\kappa_{\nu}$ (where $\kappa_{1000GHz}=10$ g cm$^{-2}$ and $\kappa_{\nu}\propto\nu^\beta$ with $\beta$=1). The uncertainty on this parameter is 10\%, following the flux calibration uncertainty of ALMA. Furthermore, the assumption of optically thin emission may not be fully valid in protoplanetary disks, and the dust mass values are technically lower limits to the pebble mass \citep{Zhu2019}. However, as all disks are computed in the same way, the assumption is sufficient for the comparison in this work.

Second, we estimate the warm dust mass in each disk using the spatially resolved ALMA emission inside the 30 K radius, where at least the CO ice is fully sublimated in the midplane, in a similar way as \citet{vanderMarel2021-c2h}. The midplane temperature profile of each disk is estimated using a power-law profile assuming a passively heated disk, following \citet{Chiang1997}:
\begin{equation}
\label{eqn:temp}
    T(r)=\left(\frac{L_*\phi}{8\pi\sigma_{SB}r^2}\right)^{1/4}
\end{equation}
with the stellar luminosity $L_*$, flaring angle $\phi$=0.02, $\sigma_{SB}$ the Stefan-Boltzmann constant and radius $r$. Transition disks generally have a different temperature profile due to their empty cavity and heated cavity edge, and their temperature profiles are estimated individually using a RADMC-3D model fitting the Spectral Energy Distribution (see Section \ref{sec:SEDs}). 

\begin{figure*}[!ht]
    \centering
    \includegraphics[width=0.7\textwidth]{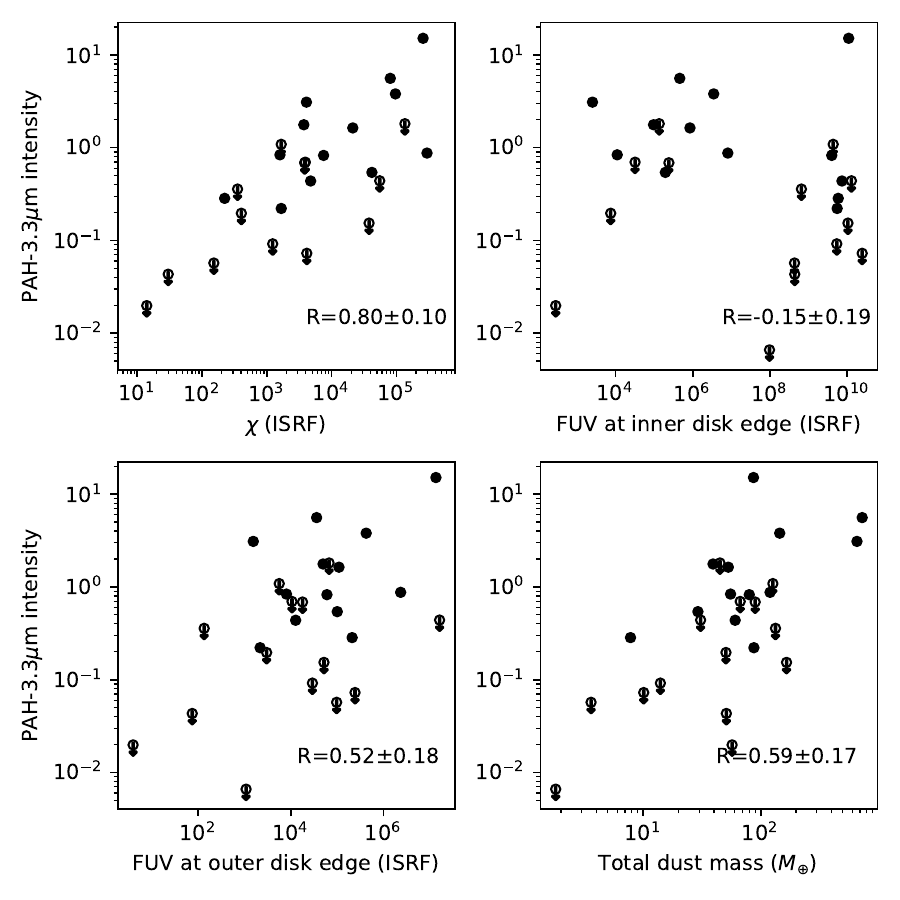}
    \caption{The 3.3 $\mu$m PAH intensity as function of various stellar and disk parameters: the parameter $\chi$ (the FUV radiation field at 150 au from the star), the FUV radiation field at the inner disk edge $R_{in}$ and at the disk outer edge $R_{size}$. The fourth plot shows the 3.3 $\mu$m PAH intensity as function of the total disk dust mass. Upper limits are indicated as empty circles with arrows. The Pearson correlation coefficient is indicated in the corner.}
    \label{fig:PAH_other}
\end{figure*}

Using these temperature profiles, we derive the radius of the CO snowline in each disk, for which different values have been found, ranging between 18 to 35 K \citep[e.g.][]{Bisschop2006,Pinte2018,Qi2019,Minissale2022,Guilloteau2025}. The warm dust mass flux is computed for each disk by integrating the millimeter flux from 0 out to the radius where the midplane temperature drops below 30 K, which is spatially resolved in all cases except WX Cha and Sz73. Figure \ref{fig:profiles} shows the radial intensity profiles of each ALMA millimeter image with the 22-30 K regime and spatial resolution indicated. Whereas the exact cutoff temperature will impact the warm dust mass, this range shows that the impact is generally negligible. Using this flux and Equation \ref{eqn:mdust} for T=30 K, we compute the warm dust mass, which is listed in Table \ref{tbl:almadata}. Although the temperature increases to $\gg$30 K within the warm dust region, the dust mass will be dominated by the temperature at the outer radius due to the shape of the temperature profile and 30 K is a reasonable choice. The uncertainty on the warm dust mass is again 10\%, following the flux calibration uncertainty of ALMA. 
For WX~Cha and Sz73, the spatial resolution is insufficient to resolve the emission around the CO snowline and we use an upper limit to the total dust mass as value.

In 10 out of 26 targets, the millimeter emission is entirely inside the 30 K radius, and the integrated flux for the warm dust mass is equal to the total flux. In that case, the difference between total and warm dust mass is caused by the assumption of the temperature in $B_{\nu}(T)$ in Eqn \ref{eqn:mdust}, and changes by a factor 1.7. For all other disks, the warm dust mass is (much) lower than the total dust mass, up to a factor 18.

\begin{figure}[!ht]
    \centering
    \includegraphics[width=0.45\textwidth]{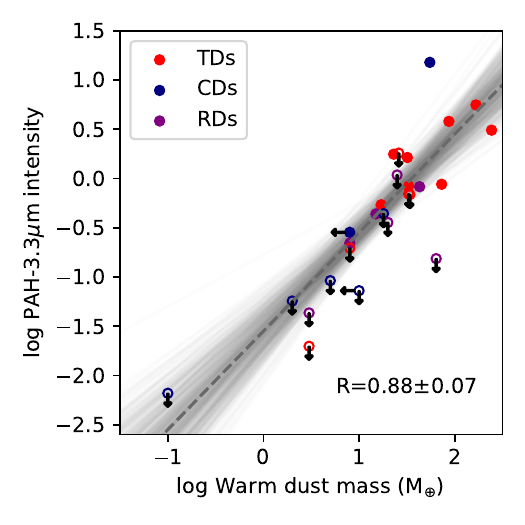}
    \caption{The 3.3 $\mu$m PAH intensity as function of the warm dust mass, computed from the millimeter flux inside the 30 K radius (CO snowline). The different colors represent the transition disks (red, TDs), compact disks (navy, CDs) and ring disks (purple, RDs). Upper limits are indicated as empty circles with arrows. The dashed line shows the best linear fit for the data points with the grey lines showing the spread between the fits. The fit has a Pearson coefficient of $r=0.88\pm0.07$ and the data are strongly correlated.}
    \label{fig:PAHmain}
\end{figure}

\section{Results}
\label{sec:results}
Figure \ref{fig:PAH_other} shows the 3.3 $\mu$m PAH intensity as a function of various radiation properties as well as the total dust mass. Using the \texttt{linmix} package \citep{Kelly2007}, we perform a linear fit to the logarithmic data including the upper limits. We have scaled the 5$\sigma$ limits on the PAH intensity to 3$\sigma$ limits for consistency with the linmix methodology in these plots and calculations. We find Pearson correlation coefficients as reported in the lower left corner of each plot: $R=0.80\pm0.10, -0.15\pm0.19, 0.52\pm0.18$ and 0.59$\pm$0.17 for correlations with the FUV radiation field $\chi$, FUV at the inner disk edge, FUV at the outer disk edge and total dust mass, respectively. We also compute the Pearson correlation coefficients of the fit of the data of the detected PAH intensities only (excluding the upper limits from the fit), which generally results in somewhat lower correlation coefficients with larger error bars: $R=0.63\pm0.23, -0.23\pm0.30, 0.41\pm0.30$ and $0.54\pm0.26$.

These results imply that a correlation exists for $\chi$, as well as a tentative correlation for the total dust mass and an even more moderate correlation with the FUV at the outer disk edge. This is consistent with earlier suggestions by models that the PAH intensity may scale with the radiation field \citep{Habart2004}. All plots show large scatter, in particular when considering the upper limits. The correlation is not unambiguous and does not explain why disks with similar spectral type or similar disk morphology show such a wide spread in PAH intensity.

Second, Figure \ref{fig:PAHmain} shows the 3.3 $\mu$m PAH intensity as function of the warm dust mass, computed from the flux inside the 30 K radius. For some disks the warm dust mass is an upper limit, as the continuum emission was insufficiently resolved with respect to the CO snowline (see Figure \ref{fig:profiles}). The data points show clear evidence for a correlation between the two parameters. Analysis with \texttt{linmix} including the PAH intensity upper limits results in a Pearson correlation coefficient of $r=0.88\pm0.07$, consistent with a very strong correlation. When considering only the data points of detected PAH 3.3 $\mu$m intensities, the Pearson correlation coefficient is $R=0.71\pm0.21$, showing that the upper limits are important drivers of the correlation.

The three groups of disks (transition, compact or ring disks) are color-coded to check for trends as well. Transition disks are found across the range of PAH intensity, consistent with a variation of warm and cold dust traps \citep{vanderMarel2023} and some of the strongest PAH intensities are found in transition disks. On the other hand, if the dust trap is `cold' (outside the 30 K radius) such as SY Cha and T Cha, the PAH intensity is much lower. Although compact disks are warm, their dust mass tends to be lower in most cases due to their smaller size and thus millimeter flux. An obvious outlier is HD179218 with a very strong PAH flux for its warm dust mass. 
Half of the ring disks show a moderate PAH intensity. Whereas ring disks have similar total dust masses as transition disks, the bulk of their dust mass is relatively cold. One outlier here is HD163296, with a rather low PAH intensity limit compared to its warm dust mass. 

\begin{figure}[!ht]
    \includegraphics[width=0.5\textwidth]{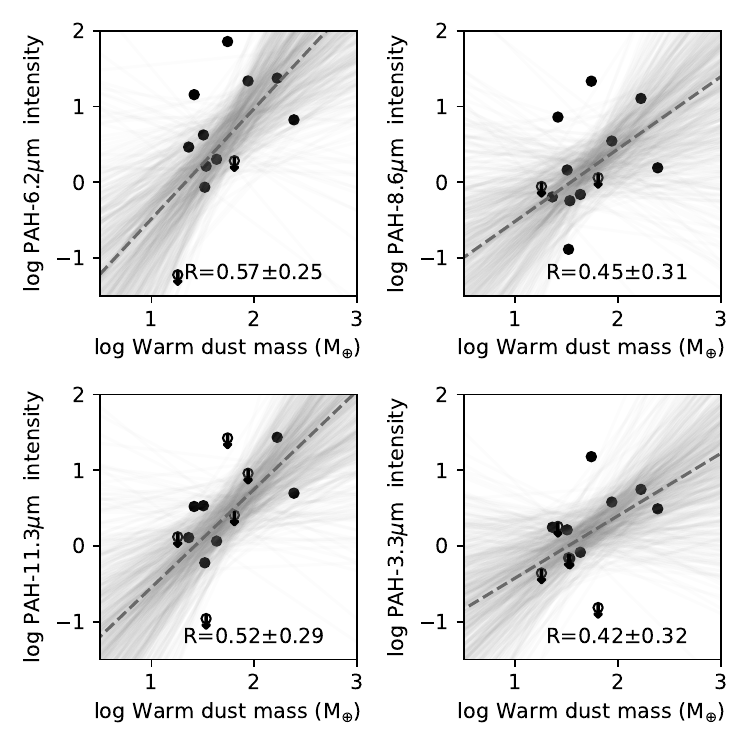}
    \caption{PAH intensity as function of warm dust mass for the long wavelength features at 6.2, 8.6 and 11.3 $\mu$m, as well as the 3.3 $\mu$m for the subsample for which the other PAH features were available (Table \ref{tab:pahacke}).}
    \label{fig:pahacke}
\end{figure}

Finally, we consider the correlation between the warm dust mass and three other PAH features at 6.2, 8.6 and 11.3 $\mu$m for a subsample of our targets. The relation between these parameters is shown in Figure \ref{fig:pahacke}. The correlation coefficient is again computed using \texttt{linmix} (including upper limits), resulting in Pearson coefficients of typically 0.5$\pm$0.3 (see Figure), i.e. moderately positive correlations, although within uncertainties still matching with the 3.3 um correlation in a much larger sample. However, if we compute the correlation coefficient for the 3.3 $\mu$m feature for this subsample only, the Pearson correlation coefficient is  $R=0.42\pm0.32$, i.e. the correlation disappears. This implies that there is some intrinsic scatter in the PAH intensities when restricting to a smaller subsample.

\section{Discussion}
\label{sec:discussion}
\subsection{Origin of PAH emission in protoplanetary disks}
The strong correlation between the warm dust mass and the 3.3$\mu$m PAH intensity regardless of disk morphology provides evidence for a scenario where PAHs are sublimated from icy dust pebbles transported vertically upwards, after those pebbles have drifted inwards toward the pressure bump(s) in the disks. This is comparable to the findings of e.g. \citet{vanderMarel2021-irs48} and \citet{Booth2021-hd100} on the release of CH$_3$OH ice into the gas-phase in IRS48 and HD100546 from the main dust traps in those disks. 
Fragmentation and vertical transport through turbulence is required, as PAHs sublimate at high temperatures $>$150-300 K compared to the midplane temperature, which are only reached at higher scale heights at the dust locations \citep{Bruderer2012,vanderMarel2021-irs48}. This is consistent with the predictions from PAH disk models from \citet{Lange2023} who have calculated in which layers of the disk PAHs are dissociated and form clusters. Our proposed scenario is similar to this idea, with the additional insight that locked up PAHs are transported with the pebbles toward their final location in the disk, where they only desorb if the local temperature is above the PAH sublimation temperature. If the majority of PAHs remain locked up in `cold' dust traps (such as seen in e.g. SY Cha and in the ring disks), the observed PAH intensity from gas-phase PAHs remains low. In this way, PAH intensities can be explained for the range of T Tauris and Herbigs as well as their individual dust substructures.

On the other hand, for PAH features at longer wavelengths these correlations are potentially weaker (Figure \ref{fig:pahacke}). This correlation only cover a small part of the sample though, including only Herbig disks. This results in larger uncertainties on the correlation coefficient, and a lack of disks where the bulk of the dust mass is `cold' (i.e. outside the CO snowline). The correlation coefficient of the 3.3 $\mu$m feature is indeed much lower as well if only this subsample is considered. Furthermore, \citet{Geers2007} reports the non-detection of the 8.6 $\mu$m PAH feature in T Cha, SY Cha, and WX Cha,  with very weak and non-detections of the 11.3 $\mu$m PAH feature, which are all T Tauri disks with small warm dust masses. As quantitative information is missing, these targets could not be added to our plot, but such data certainly strengthens the correlation. The observed trend at the long wavelength PAH features thus does not contradict the findings for the 3.3 $\mu$m feature and is consistent with the main hypothesis. It is also possible that the PAH intensity is not a sufficiently direct tracer of the amount of released PAHs, and thus, of the amount of warm dust pebbles, and therefore, the correlation is not reflected by the properties of the detected targets alone: the correlation is driven by the combination of deep upper limits and strong detections corresponding to a range of warm dust masses.

If PAHs are indeed locked up in ices in protoplanetary disks and only revealed in infrared observations due to sublimation, this implicates significantly high amounts of PAHs being present in planet forming regions, but mostly in solid form. This is particularly interesting in high UV regions, as irradiated ices have been shown to rapidly form organic macromolecules \citep{Ligterink2024}. The destruction of PAHs through UV radiation into smaller components may result in a more efficient formation process of such macromolecules \citep{Alexander2008}. On the other hand, there is also the possibility that (frozen) PAHs in disks are not inherited from the interstellar medium, but formed in situ under the influence of radiation. However, the omnipresence of PAHs in the ISM makes this unnecessary, and a thorough investigation of the fractional abundances of PAHs in disks compared to ISM would be required to test this.

\subsection{Outliers in the warm dust - PAH intensity correlation}
The two exceptions in the main correlation with the 3.3 $\mu$m feature in Figure \ref{fig:PAHmain} are HD179218 and HD163296. HD179218 has a very high PAH intensity for its warm dust mass, also seen in the correlations with the 6.2 and 8.6 $\mu$m features. However, since this is a compact disk, it is plausible that the warm dust mass is  underestimated by a factor of a few due to optically thick emission, as Eqn. \ref{eqn:mdust} is only valid for optically thin emission. This would bring this datapoint in line with the observed trend. Second, HD163296 appears to lie below the trend with an upper limit on the 3.3 $\mu$m PAH intensity and a fairly high warm dust mass. However, inspection of its radial profile in Figure \ref{fig:profiles} shows that there is a rather wide dust gap outside the CO snowline and it is possible that the majority of icy dust pebbles is halted outside, lowering the amount that ends up in the warm region. This would decrease the amount of sublimating icy pebbles in this disk and move also this datapoint closer to the general trend. This is a complex problem as it depends on which dust traps form first in the disk \citep{Pinilla2015}. Quantifying the amount of icy pebbles that make it through the gap is however beyond the scope of this work. Removing HD179218 and HD163296 from the sample of Figure \ref{fig:PAHmain} increases the Pearson correlation coefficient to $R=0.94\pm0.04$. Potential other outliers in the long wavelength PAH features in Figure \ref{fig:pahacke} are HD104237 and HD139614, for which the PAH upper limits appear to be very low compared to the main trend.

In the sample selection, a number of targets from the original samples of \citet{Habart2004} and \citet{Geers2007} were excluded to ensure that the analysis would be consistent. For the majority of those targets only upper limits on the PAH 3.3 $\mu$m were reported. A remarkably bright PAH source is HD98922, a massive A0 Herbig star at 650 pc of 5.5 $M_{\odot}$ and 1200 $L_{\odot}$ with a flat SED and faint polarized emission \citep{Garufi2022}. As these authors report, only low-resolution ALMA data exist (2015.1.01600.S) and the disk is detected but unresolved. If this system follows the trend from Figure \ref{fig:PAHmain}, the warm dust mass is expected to be very high. Determining the warm dust mass would require higher resolution data and a full modeling of the temperature structure which is left for future work. 

\subsection{Spatially resolved PAH emission}
\begin{figure}[!ht]
    \centering
    \includegraphics[width=0.5\textwidth]{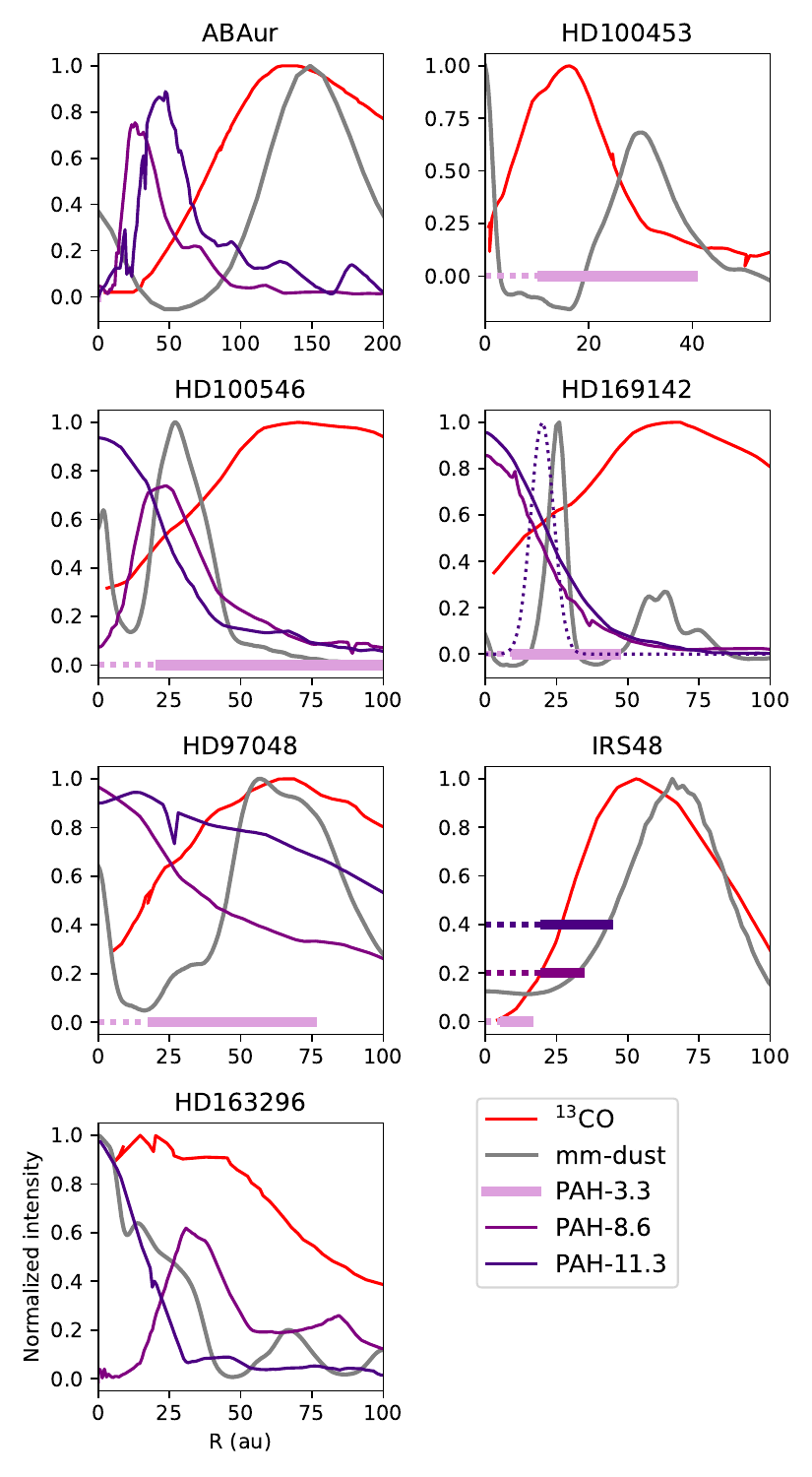}
    \caption{Radial profiles of the $^{13}$CO 3-2 or 2-1 line (blue) and ALMA continuum (gray), in combination with the maximum spatial extent of the PAH 3.3 $\mu$m emission (plum bar), and the derived spatial profiles of the PAH 8.6 (purple) and PAH 11.3 (indigo) features when available. The spatial extent is not constrained for the inner edge of the disk, so the inner region (PSF radius) is indicated as dotted rather than solid in the horizontal bar. For IRS48, only spatial extent is available for the 8.6 and 11.3 features. For HD169142, the 11.3 emission is consistent with a ring \citep{Devinat2022,Yoffe2023}, which is overplotted as a dotted line. The $^{13}$CO curves are taken from \citet{Wolfer2023} and \citet{Zhang2021}, the continuum curves from this work. For HD100453 $^{12}$CO 3-2 instead of $^{13}$CO 3-2 is plotted, as those are the only available line data for this source. The spatial extent of the PAH emission is generally just inside the dust cavity.}
    \label{fig:spatial}
\end{figure}

From the integrated fluxes studied in this work it remains unclear if the PAH emission indeed originates from the dust trap region. For a subset of 7 targets in our sample (AB Aur, HD100453, HD100546, HD169142, HD97048, IRS48 and HD163206, mostly transition disks), spatially resolved information of the PAH features is available in the literature. The spatial extent of the 3.3 $\mu$m feature has been estimated for HD97048, IRS48, HD100453, HD100546 and HD169142 using VLT/NACO data \citep{Habart2006,Geers2007,Bouteraon2019}. These profiles are somewhat broader than the PSF of $\sim$0.1-0.2" so although it is not possible to radially resolve a profile or constrain the inner edge, the spatial extent could be determined, which ranges up to radii of 0.9". For the 8.6 and 11.3 $\mu$m feature, the spatially resolved intensity profiles have been obtained with the VISIR-NEAR instrument for ABAur, HD100546, HD97048, HD169142 and HD163296 \citep{Yoffe2023}. For IRS48, the spatial extent of these features is constrained with VISIR where the inner edge remains unconstrained \citep{Geers2007-irs48}.

Figure \ref{fig:spatial} shows the spatial extent of the 3.3, 8.6 and 11.3 $\mu$m features from the literature described above. The 3.3 $\mu$m feature is indicated as a horizontal bar, where the inner part shows a dotted profile of the PSF size to indicate the inner edge is unconstrained. Since the 3.3 $\mu$m feature is only marginally resolved, its emission may be ring-like as well. The other two features are taken directly from \citep{Yoffe2023}, where available. For HD169142, the 11.3 $\mu$m emission is consistent with a ring of 0.14-0.17" radius \citep{Devinat2022,Yoffe2023}, so this is overlaid as well. These PAH extents are compared in Figure \ref{fig:spatial} with the $^{13}$CO 3--2 or 2--1 radial profiles taken from \citet{Wolfer2023} and the ALMA continuum as derived in Figure \ref{fig:profiles}. For HD100453 only $^{12}$CO 3-2 data were available. The spatially resolved $^{13}$CO 3--2 or 2--1 line provides a proxy of the gas gap inside the dust cavity \citep{vanderMarel2016-isot,Facchini2017}, although the gap depth could be overestimated if the gas is too warm to excite these transitions \citep{Leemker2022}. The $^{12}$CO line of HD100453 becomes optically thin as well in the very inner part of the cavity, revealing the deepest part of the gas gap.

For AB~Aur, HD100546, HD169142, HD97048, IRS48 and potentially HD100453 the PAH emission peaks inside the cavity. For HD169142 and HD100546, the spatially resolved PAH ring is close to the inner edge of the dust ring, for AB Aur the PAH ring is well inside. For HD163296, the PAH ring appears to coincide with the third dust ring. Therefore, the spatial information also shows that PAH emission is possibly correlated with the dust trap, hinting that these could be PAHs sublimating from vertically stirred grains. The emission does not fully correspond to the location of the dust ring or dust crescent itself like for example the CH$_3$OH emission in IRS48 \citep{vanderMarel2021-irs48}. However, since gas extends within the cavity as seen in the $^{13}$CO emission, as well as from the high accretion rates of each of these disks (Table \ref{tbl:sample}), the sublimated PAHs may  follow the accretion flow through the gap without getting destroyed, and emit primarily in the inner region with the highest UV radiation. In contrast, the reason that the CH$_3$OH emission in IRS48 has the same morphology as the dust crescent is because the gas molecules are rapidly photodissociated outside the dust trap, and continuously replenished at the dust trap through sublimation \citep{Temmink2025}. Furthermore, the PAH emission sometimes also extends out to larger radii. Additional radial transport, such as turbulent mixing and diffusion likely play a role here, and sublimation is not the sole process setting the radial extent of the PAH emission. Further studies including sublimation and transport processes are required to fully quantify this.

\section{Conclusions}
In this study, we analyzed the PAH intensity of the 3.3 $\mu$m feature for a large sample of T Tauri and Herbig disks from the literature, as well as the 6.2, 8.6 and 11.3 $\mu$m PAH features for a subset of Herbig disks, to test for correlations with disk and radiation properties. We conclude the following:
\begin{itemize}
    \item A strong correlation with a Pearson coefficient of 0.88$\pm$0.08 was found between the 3.3 $\mu$m PAH intensity and the warm dust mass, which is defined as the millimeter dust mass measured from ALMA continuum observations within the 30 K midplane radius.
    \item The 3.3 $\mu$m PAH intensity has (much) weaker correlations with other relevant parameters such as the FUV luminosity, the FUV radiation field at inner and outer edge of the disk and the total dust mass.
    \item The 6.2, 8.6 and 11.3 $\mu$m features show somewhat weaker correlations with warm dust mass for a smaller subsample, but within uncertainties matching with the correlation of the 3.3 um feature. 
    \item The spatial distribution of the PAH features (where available) is generally  inside of the dust trap locations, suggesting that gas-phase PAHs may follow the accretion flow after sublimation without getting destroyed, although additional transport like diffusion and turbulent mixing may also play a role.    
    \item The correlation with warm dust mass is consistent with the hypothesis that PAHs are generally frozen out in protoplanetary disks, and only become detectable in the gas-phase when icy pebbles drift inwards, get transported vertically upwards and sublimate their icy layers, similar to previous findings for complex organic molecules (COMs) in dust traps \citep{vanderMarel2021-irs48,Booth2021-hd100}. This can explain the PAH intensity w.r.t. the range in dust disk morphologies and stellar luminosities in the sample.
    \item The findings regarding PAH sublimation provide further evidence that disk composition is governed by icy pebble transport, rather than a static chemical equilibrium disk.
\end{itemize}

\begin{acknowledgements} 
We thank the referee for their thorough report and useful suggestions. We would also like to thank Margot Leemker, Kevin Lange, Rens Waters, Ella de Vries and Judith Moulijn for useful discussions. This paper makes use of the following ALMA data: ADS/JAO.ALMA\#2012.1.00631.S,
2013.1.00100.S,
2013.1.00157.S,
2013.1.00498.S,
2015.1.00773.S,
2015.1.00979.S,
2016.1.00336.S,
2016.1.00340.S,
2016.1.00344.S,
2016.1.00484.L,
2016.1.00826.S,
2017.1.01424.S,
2018.1.00689.S,
2018.1.01458.S,
2021.1.00709.S,
2021.1.00854.S,
2021.1.01137.S,
2022.1.01302.S. 
ALMA is a partnership of ESO (representing its member states), NSF (USA) and NINS (Japan), together with NRC (Canada), NSTC and ASIAA (Taiwan), and KASI (Republic of Korea), in cooperation with the Republic of Chile. The Joint ALMA Observatory is operated by ESO, AUI/NRAO and NAOJ.
\end{acknowledgements}

\bibliographystyle{aa}
\bibliography{refs.bib}

@ARTICLE{Allamandola1985,
       author = {{Allamandola}, L.~J. and {Tielens}, A.~G.~G.~M. and {Barker}, J.~R.},
        title = "{Polycyclic aromatic hydrocarbons and the unidentified infrared emission bands: auto exhaust along the milky way.}",
      journal = {\apjl},
         year = 1985,
        month = mar,
       volume = {290},
        pages = {L25-L28},
          doi = {10.1086/184435},
       adsurl = {https://ui.adsabs.harvard.edu/abs/1985ApJ...290L..25A}
}

@ARTICLE{Visser2007,
       author = {{Visser}, R. and {Geers}, V.~C. and {Dullemond}, C.~P. and {Augereau}, J. -C. and {Pontoppidan}, K.~M. and {van Dishoeck}, E.~F.},
        title = "{PAH chemistry and IR emission from circumstellar disks}",
      journal = {\aap},
         year = 2007,
        month = apr,
       volume = {466},
       number = {1},
        pages = {229-241},
          doi = {10.1051/0004-6361:20066829},
archivePrefix = {arXiv},
       eprint = {astro-ph/0701606},
 primaryClass = {astro-ph},
       adsurl = {https://ui.adsabs.harvard.edu/abs/2007A&A...466..229V}
}

@ARTICLE{Peeters2004,
       author = {{Peeters}, E. and {Spoon}, H.~W.~W. and {Tielens}, A.~G.~G.~M.},
        title = "{Polycyclic Aromatic Hydrocarbons as a Tracer of Star Formation?}",
      journal = {\apj},
         year = 2004,
        month = oct,
       volume = {613},
       number = {2},
        pages = {986-1003},
          doi = {10.1086/423237},
archivePrefix = {arXiv},
       eprint = {astro-ph/0406183},
 primaryClass = {astro-ph},
       adsurl = {https://ui.adsabs.harvard.edu/abs/2004ApJ...613..986P}
}

@ARTICLE{Geers2007,
       author = {{Geers}, V.~C. and {van Dishoeck}, E.~F. and {Visser}, R. and {Pontoppidan}, K.~M. and {Augereau}, J. -C. and {Habart}, E. and {Lagrange}, A.~M.},
        title = "{Spatially extended polycyclic aromatic hydrocarbons in circumstellar disks around T Tauri and Herbig Ae stars}",
      journal = {\aap},
         year = 2007,
        month = dec,
       volume = {476},
       number = {1},
        pages = {279-289},
          doi = {10.1051/0004-6361:20078466},
archivePrefix = {arXiv},
       eprint = {0710.2825},
 primaryClass = {astro-ph},
       adsurl = {https://ui.adsabs.harvard.edu/abs/2007A&A...476..279G}
}

@ARTICLE{Habart2006,
       author = {{Habart}, E. and {Natta}, A. and {Testi}, L. and {Carbillet}, M.},
        title = "{Spatially resolved PAH emission in the inner disks of Herbig Ae/Be stars}",
      journal = {\aap},
         year = 2006,
        month = apr,
       volume = {449},
       number = {3},
        pages = {1067-1075},
          doi = {10.1051/0004-6361:20052994},
archivePrefix = {arXiv},
       eprint = {astro-ph/0503105},
 primaryClass = {astro-ph},
       adsurl = {https://ui.adsabs.harvard.edu/abs/2006A&A...449.1067H}
}

@ARTICLE{Habart2004,
       author = {{Habart}, E. and {Natta}, A. and {Kr{\"u}gel}, E.},
        title = "{PAHs in circumstellar disks around Herbig Ae/Be stars}",
      journal = {\aap},
         year = 2004,
        month = nov,
       volume = {427},
        pages = {179-192},
          doi = {10.1051/0004-6361:20035916},
archivePrefix = {arXiv},
       eprint = {astro-ph/0405195},
 primaryClass = {astro-ph},
       adsurl = {https://ui.adsabs.harvard.edu/abs/2004A&A...427..179H}
}

@ARTICLE{Acke2010,
       author = {{Acke}, B. and {Bouwman}, J. and {Juh{\'a}sz}, A. and {Henning}, Th. and {van den Ancker}, M.~E. and {Meeus}, G. and {Tielens}, A.~G.~G.~M. and {Waters}, L.~B.~F.~M.},
        title = "{Spitzer's View on Aromatic and Aliphatic Hydrocarbon Emission in Herbig Ae Stars}",
      journal = {\apj},
         year = 2010,
        month = jul,
       volume = {718},
       number = {1},
        pages = {558-574},
          doi = {10.1088/0004-637X/718/1/558},
archivePrefix = {arXiv},
       eprint = {1006.1130},
 primaryClass = {astro-ph.SR},
       adsurl = {https://ui.adsabs.harvard.edu/abs/2010ApJ...718..558A}
}

@ARTICLE{Meeus2001,
   author = {{Meeus}, G. and {Waters}, L.~B.~F.~M. and {Bouwman}, J. and 
	{van den Ancker}, M.~E. and {Waelkens}, C. and {Malfait}, K.
	},
    title = "{ISO spectroscopy of circumstellar dust in 14 Herbig Ae/Be systems: Towards an understanding of dust processing}",
  journal = {\aap},
   eprint = {astro-ph/0012295},
     year = 2001,
    month = jan,
   volume = 365,
    pages = {476-490},
      doi = {10.1051/0004-6361:20000144},
   adsurl = {http://adsabs.harvard.edu/abs/2001A%26A...365..476M}
}

@ARTICLE{Siebenmorgen2010,
       author = {{Siebenmorgen}, R. and {Kr{\"u}gel}, E.},
        title = "{The destruction and survival of polycyclic aromatic hydrocarbons in the disks of T Tauri stars}",
      journal = {\aap},
         year = 2010,
        month = feb,
       volume = {511},
          eid = {A6},
        pages = {A6},
          doi = {10.1051/0004-6361/200912035},
       adsurl = {https://ui.adsabs.harvard.edu/abs/2010A&A...511A...6S}
}

@ARTICLE{Geers2009,
       author = {{Geers}, V.~C. and {van Dishoeck}, E.~F. and {Pontoppidan}, K.~M. and {Lahuis}, F. and {Crapsi}, A. and {Dullemond}, C.~P. and {Blake}, G.~A.},
        title = "{Lack of PAH emission toward low-mass embedded young stellar objects}",
      journal = {\aap},
         year = 2009,
        month = mar,
       volume = {495},
       number = {3},
        pages = {837-846},
          doi = {10.1051/0004-6361:200811001},
archivePrefix = {arXiv},
       eprint = {0812.3664},
 primaryClass = {astro-ph},
       adsurl = {https://ui.adsabs.harvard.edu/abs/2009A&A...495..837G}
}

@ARTICLE{Bouwman2010,
       author = {{Bouwman}, J. and {Cuppen}, H.~M. and {Bakker}, A. and {Allamandola}, L.~J. and {Linnartz}, H.},
        title = "{Photochemistry of the PAH pyrene in water ice: the case for ion-mediated solid-state astrochemistry}",
      journal = {\aap},
         year = 2010,
        month = feb,
       volume = {511},
          eid = {A33},
        pages = {A33},
          doi = {10.1051/0004-6361/200913291},
archivePrefix = {arXiv},
       eprint = {0911.1750},
 primaryClass = {astro-ph.SR},
       adsurl = {https://ui.adsabs.harvard.edu/abs/2010A&A...511A..33B}
}

@ARTICLE{Rapaciola2009,
       author = {{Rapacioli}, M. and {Spiegelman}, F.},
        title = "{Modelling singly ionized coronene clusters}",
      journal = {European Physical Journal D},
         year = 2009,
        month = apr,
       volume = {52},
       number = {1-3},
        pages = {55-58},
          doi = {10.1140/epjd/e2008-00280-2},
       adsurl = {https://ui.adsabs.harvard.edu/abs/2009EPJD...52...55R}
}

@ARTICLE{Lange2023,
       author = {{Lange}, K. and {Dominik}, C. and {Tielens}, A.~G.~G.~M.},
        title = "{Turbulent processing of PAHs in protoplanetary discs. Coagulation and freeze-out leading to depletion of gas-phase PAHs}",
      journal = {\aap},
         year = 2023,
        month = jun,
       volume = {674},
          eid = {A200},
        pages = {A200},
          doi = {10.1051/0004-6361/202245108},
archivePrefix = {arXiv},
       eprint = {2303.07981},
 primaryClass = {astro-ph.EP},
       adsurl = {https://ui.adsabs.harvard.edu/abs/2023A&A...674A.200L}
}

@ARTICLE{Lange2021,
       author = {{Lange}, K. and {Dominik}, C. and {Tielens}, A.~G.~G.~M.},
        title = "{Stability of polycyclic aromatic hydrocarbon clusters in protoplanetary discs}",
      journal = {\aap},
         year = 2021,
        month = sep,
       volume = {653},
          eid = {A21},
        pages = {A21},
          doi = {10.1051/0004-6361/202140590},
archivePrefix = {arXiv},
       eprint = {2108.10695},
 primaryClass = {astro-ph.EP},
       adsurl = {https://ui.adsabs.harvard.edu/abs/2021A&A...653A..21L}
}

@ARTICLE{Andrews2020,
       author = {{Andrews}, Sean M.},
        title = "{Observations of Protoplanetary Disk Structures}",
      journal = {\araa},
         year = 2020,
        month = aug,
       volume = {58},
        pages = {483-528},
          doi = {10.1146/annurev-astro-031220-010302},
archivePrefix = {arXiv},
       eprint = {2001.05007},
 primaryClass = {astro-ph.EP},
       adsurl = {https://ui.adsabs.harvard.edu/abs/2020ARA&A..58..483A}
}

@ARTICLE{Pinilla2012b,
   author = {{Pinilla}, P. and {Benisty}, M. and {Birnstiel}, T.},
    title = "{Ring shaped dust accumulation in transition disks}",
  journal = {\aap},
archivePrefix = "arXiv",
   eprint = {1207.6485},
 primaryClass = "astro-ph.EP",
     year = 2012,
    month = sep,
   volume = 545,
      eid = {A81},
    pages = {A81},
      doi = {10.1051/0004-6361/201219315},
   adsurl = {http://adsabs.harvard.edu/abs/2012A%26A...545A..81P}
}

@ARTICLE{Booth2021-hd100,
       author = {{Booth}, Alice S. and {Walsh}, Catherine and {Terwisscha van Scheltinga}, Jeroen and {van Dishoeck}, Ewine F. and {Ilee}, John D. and {Hogerheijde}, Michiel R. and {Kama}, Mihkel and {Nomura}, Hideko},
        title = "{An inherited complex organic molecule reservoir in a warm planet-hosting disk}",
      journal = {Nature Astronomy},
         year = 2021,
        month = jan,
       volume = {5},
        pages = {684-690},
          doi = {10.1038/s41550-021-01352-w},
archivePrefix = {arXiv},
       eprint = {2104.08348},
 primaryClass = {astro-ph.EP},
       adsurl = {https://ui.adsabs.harvard.edu/abs/2021NatAs...5..684B}
}

@ARTICLE{vanderMarel2021-c2h,
       author = {{van der Marel}, Nienke and {Bosman}, Arthur D. and {Krijt}, Sebastiaan and {Mulders}, Gijs D. and {Bergner}, Jennifer B.},
        title = "{If you like C/O variations, you should have put a ring on it}",
      journal = {\aap},
         year = 2021,
        month = sep,
       volume = {653},
          eid = {L9},
        pages = {L9},
          doi = {10.1051/0004-6361/202141786},
archivePrefix = {arXiv},
       eprint = {2108.07679},
 primaryClass = {astro-ph.EP},
       adsurl = {https://ui.adsabs.harvard.edu/abs/2021A&A...653L...9V}
}

@ARTICLE{vanderMarel2021-irs48,
       author = {{van der Marel}, Nienke and {Booth}, Alice S. and {Leemker}, Margot and {van Dishoeck}, Ewine F. and {Ohashi}, Satoshi},
        title = "{A major asymmetric ice trap in a planet-forming disk. I. Formaldehyde and methanol}",
      journal = {\aap},
         year = 2021,
        month = jul,
       volume = {651},
          eid = {L5},
        pages = {L5},
          doi = {10.1051/0004-6361/202141051},
archivePrefix = {arXiv},
       eprint = {2104.08906},
 primaryClass = {astro-ph.EP},
       adsurl = {https://ui.adsabs.harvard.edu/abs/2021A&A...651L...5V}
}

@ARTICLE{Brunken2022,
       author = {{Brunken}, Nashanty G.~C. and {Booth}, Alice S. and {Leemker}, Margot and {Nazari}, Pooneh and {van der Marel}, Nienke and {van Dishoeck}, Ewine F.},
        title = "{A major asymmetric ice trap in a planet-forming disk. III. First detection of dimethyl ether}",
      journal = {\aap},
         year = 2022,
        month = mar,
       volume = {659},
          eid = {A29},
        pages = {A29},
          doi = {10.1051/0004-6361/202142981},
archivePrefix = {arXiv},
       eprint = {2203.02936},
 primaryClass = {astro-ph.EP},
       adsurl = {https://ui.adsabs.harvard.edu/abs/2022A&A...659A..29B}
}

@ARTICLE{Booth2024,
       author = {{Booth}, Alice S. and {Leemker}, Margot and {van Dishoeck}, Ewine F. and {Evans}, Lucy and {Ilee}, John D. and {Kama}, Mihkel and {Keyte}, Luke and {Law}, Charles J. and {van der Marel}, Nienke and {Nomura}, Hideko and {Notsu}, Shota and {{\"O}berg}, Karin and {Temmink}, Milou and {Walsh}, Catherine},
        title = "{An ALMA Molecular Inventory of Warm Herbig Ae Disks. I. Molecular Rings, Asymmetries, and Complexity in the HD 100546 Disk}",
      journal = {\aj},
         year = 2024,
        month = apr,
       volume = {167},
       number = {4},
          eid = {164},
        pages = {164},
          doi = {10.3847/1538-3881/ad2700},
archivePrefix = {arXiv},
       eprint = {2402.04001},
 primaryClass = {astro-ph.EP},
       adsurl = {https://ui.adsabs.harvard.edu/abs/2024AJ....167..164B}
}

@ARTICLE{vanderMarel2023,
       author = {{van der Marel}, Nienke},
        title = "{Transition disks: the observational revolution from SEDs to imaging}",
      journal = {European Physical Journal Plus},
         year = 2023,
        month = mar,
       volume = {138},
       number = {3},
          eid = {225},
        pages = {225},
          doi = {10.1140/epjp/s13360-022-03628-0},
archivePrefix = {arXiv},
       eprint = {2210.05539},
 primaryClass = {astro-ph.EP},
       adsurl = {https://ui.adsabs.harvard.edu/abs/2023EPJP..138..225V}
}

@ARTICLE{Temmink2023,
       author = {{Temmink}, M. and {Booth}, A.~S. and {van der Marel}, N. and {van Dishoeck}, E.~F.},
        title = "{Investigating the asymmetric chemistry in the disk around the young star HD 142527}",
      journal = {\aap},
         year = 2023,
        month = jul,
       volume = {675},
          eid = {A131},
        pages = {A131},
          doi = {10.1051/0004-6361/202346272},
archivePrefix = {arXiv},
       eprint = {2304.06382},
 primaryClass = {astro-ph.EP},
       adsurl = {https://ui.adsabs.harvard.edu/abs/2023A&A...675A.131T}
}

@ARTICLE{Ligterink2024,
       author = {{Ligterink}, Niels F.~W. and {Pinilla}, Paola and {van der Marel}, Nienke and {van Scheltinga}, Jeroen Terwisscha and {Booth}, Alice S. and {Alexander}, Conel M. O'D. and {Riebe}, My E.~I.},
        title = "{The rapid formation of macromolecules in irradiated ice of protoplanetary disk dust traps}",
      journal = {Nature Astronomy},
         year = 2024,
        month = jul,
          doi = {10.1038/s41550-024-02334-4},
       adsurl = {https://ui.adsabs.harvard.edu/abs/2024NatAs.tmp..142L}
}

@ARTICLE{Zhang2021,
       author = {{Zhang}, Ke and {Booth}, Alice S. and {Law}, Charles J. and {Bosman}, Arthur D. and {Schwarz}, Kamber R. and {Bergin}, Edwin A. and {{\"O}berg}, Karin I. and {Andrews}, Sean M. and {Guzm{\'a}n}, Viviana V. and {Walsh}, Catherine and {Qi}, Chunhua and {van't Hoff}, Merel L.~R. and {Long}, Feng and {Wilner}, David J. and {Huang}, Jane and {Czekala}, Ian and {Ilee}, John D. and {Cataldi}, Gianni and {Bergner}, Jennifer B. and {Aikawa}, Yuri and {Teague}, Richard and {Bae}, Jaehan and {Loomis}, Ryan A. and {Calahan}, Jenny K. and {Alarc{\'o}n}, Felipe and {M{\'e}nard}, Fran{\c{c}}ois and {Le Gal}, Romane and {Sierra}, Anibal and {Yamato}, Yoshihide and {Nomura}, Hideko and {Tsukagoshi}, Takashi and {P{\'e}rez}, Laura M. and {Trapman}, Leon and {Liu}, Yao and {Furuya}, Kenji},
        title = "{Molecules with ALMA at Planet-forming Scales (MAPS). V. CO Gas Distributions}",
      journal = {\apjs},
         year = 2021,
        month = nov,
       volume = {257},
       number = {1},
          eid = {5},
        pages = {5},
          doi = {10.3847/1538-4365/ac1580},
archivePrefix = {arXiv},
       eprint = {2109.06233},
 primaryClass = {astro-ph.EP},
       adsurl = {https://ui.adsabs.harvard.edu/abs/2021ApJS..257....5Z}
}

@ARTICLE{Temmink2025,
       author = {{Temmink}, Milou and {Booth}, Alice S. and {Leemker}, Margot and {van der Marel}, Nienke and {van Dishoeck}, Ewine F. and {Evans}, Lucy and {Keyte}, Luke and {Law}, Charles J. and {Notsu}, Shota and {{\"O}berg}, Karin and {Walsh}, Catherine},
        title = "{Characterising the molecular line emission in the asymmetric Oph-IRS 48 dust trap: Temperatures, timescales, and sub-thermal excitation}",
      journal = {\aap},
         year = 2025,
        month = jan,
       volume = {693},
          eid = {A101},
        pages = {A101},
          doi = {10.1051/0004-6361/202452175},
archivePrefix = {arXiv},
       eprint = {2411.12418},
 primaryClass = {astro-ph.EP},
       adsurl = {https://ui.adsabs.harvard.edu/abs/2025A&A...693A.101T}
}

@ARTICLE{Devinat2022,
       author = {{Devinat}, Marie and {Habart}, {\'E}milie and {Pantin}, {\'E}ric and {Ysard}, Nathalie and {Jones}, Anthony and {Labadie}, Lucas and {Di Folco}, Emmanuel},
        title = "{Radial distribution of the carbonaceous nano-grains in the protoplanetary disk around HD 169142}",
      journal = {\aap},
         year = 2022,
        month = jul,
       volume = {663},
          eid = {A151},
        pages = {A151},
          doi = {10.1051/0004-6361/202243112},
archivePrefix = {arXiv},
       eprint = {2205.00213},
 primaryClass = {astro-ph.EP},
       adsurl = {https://ui.adsabs.harvard.edu/abs/2022A&A...663A.151D}
}

@ARTICLE{vdMMulders2021,
       author = {{van der Marel}, Nienke and {Mulders}, Gijs D.},
        title = "{A Stellar Mass Dependence of Structured Disks: A Possible Link with Exoplanet Demographics}",
      journal = {\aj},
         year = 2021,
        month = jul,
       volume = {162},
       number = {1},
          eid = {28},
        pages = {28},
          doi = {10.3847/1538-3881/ac0255},
archivePrefix = {arXiv},
       eprint = {2104.06838},
 primaryClass = {astro-ph.EP},
       adsurl = {https://ui.adsabs.harvard.edu/abs/2021AJ....162...28V}
}

@ARTICLE{Geers2006,
   author = {{Geers}, V.~C. and {Augereau}, J.-C. and {Pontoppidan}, K.~M. and 
	{Dullemond}, C.~P. and {Visser}, R. and {Kessler-Silacci}, J.~E. and 
	{Evans}, II, N.~J. and {van Dishoeck}, E.~F. and {Blake}, G.~A. and 
	{Boogert}, A.~C.~A. and {Brown}, J.~M. and {Lahuis}, F. and 
	{Mer{\'{\i}}n}, B.},
    title = "{C2D Spitzer-IRS spectra of disks around T Tauri stars. II. PAH emission features}",
  journal = {\aap},
   eprint = {astro-ph/0609157},
     year = 2006,
    month = nov,
   volume = 459,
    pages = {545-556},
      doi = {10.1051/0004-6361:20064830},
   adsurl = {http://adsabs.harvard.edu/abs/2006A%26A...459..545G}
}

@ARTICLE{Geers2007-irs48,
       author = {{Geers}, V.~C. and {Pontoppidan}, K.~M. and {van Dishoeck}, E.~F. and {Dullemond}, C.~P. and {Augereau}, J. -C. and {Mer{\'\i}n}, B. and {Oliveira}, I. and {Pel}, J.~W.},
        title = "{Spatial separation of small and large grains in the transitional disk around the young star <ASTROBJ>IRS 48</ASTROBJ>}",
      journal = {\aap},
         year = 2007,
        month = jul,
       volume = {469},
       number = {3},
        pages = {L35-L38},
          doi = {10.1051/0004-6361:20077524},
archivePrefix = {arXiv},
       eprint = {0705.2969},
 primaryClass = {astro-ph},
       adsurl = {https://ui.adsabs.harvard.edu/abs/2007A&A...469L..35G}
}

@ARTICLE{Habing1968,
       author = {{Habing}, H.~J.},
        title = "{The interstellar radiation density between 912 A and 2400 A}",
      journal = {\bain},
         year = 1968,
        month = jan,
       volume = {19},
        pages = {421},
       adsurl = {https://ui.adsabs.harvard.edu/abs/1968BAN....19..421H}
}

@ARTICLE{Herczeg2014,
       author = {{Herczeg}, Gregory J. and {Hillenbrand}, Lynne A.},
        title = "{An Optical Spectroscopic Study of T Tauri Stars. I. Photospheric Properties}",
      journal = {\apj},
         year = 2014,
        month = may,
       volume = {786},
       number = {2},
          eid = {97},
        pages = {97},
          doi = {10.1088/0004-637X/786/2/97},
archivePrefix = {arXiv},
       eprint = {1403.1675},
 primaryClass = {astro-ph.SR},
       adsurl = {https://ui.adsabs.harvard.edu/abs/2014ApJ...786...97H}
}

@ARTICLE{Grant2023,
       author = {{Grant}, Sierra L. and {Stapper}, Lucas M. and {Hogerheijde}, Michiel R. and {van Dishoeck}, Ewine F. and {Brittain}, Sean and {Vioque}, Miguel},
        title = "{The \textbackslashdot\{M\} -M $_{disk}$ Relationship for Herbig Ae/Be Stars: A Lifetime Problem for Disks with Low Masses?}",
      journal = {\aj},
         year = 2023,
        month = oct,
       volume = {166},
       number = {4},
          eid = {147},
        pages = {147},
          doi = {10.3847/1538-3881/acf128},
archivePrefix = {arXiv},
       eprint = {2308.11430},
 primaryClass = {astro-ph.SR},
       adsurl = {https://ui.adsabs.harvard.edu/abs/2023AJ....166..147G}
}

@INPROCEEDINGS{Manara2023,
       author = {{Manara}, C.~F. and {Ansdell}, M. and {Rosotti}, G.~P. and {Hughes}, A.~M. and {Armitage}, P.~J. and {Lodato}, G. and {Williams}, J.~P.},
        title = "{Demographics of Young Stars and their Protoplanetary Disks: Lessons Learned on Disk Evolution and its Connection to Planet Formation}",
    booktitle = {Protostars and Planets VII},
         year = 2023,
       editor = {{Inutsuka}, S. and {Aikawa}, Y. and {Muto}, T. and {Tomida}, K. and {Tamura}, M.},
       series = {Astronomical Society of the Pacific Conference Series},
       volume = {534},
        month = jul,
        pages = {539},
          doi = {10.48550/arXiv.2203.09930},
archivePrefix = {arXiv},
       eprint = {2203.09930},
 primaryClass = {astro-ph.SR},
       adsurl = {https://ui.adsabs.harvard.edu/abs/2023ASPC..534..539M}
}

@ARTICLE{Andrews2018,
   author = {{Andrews}, S.~M. and {Huang}, J. and {P{\'e}rez}, L.~M. and 
	{Isella}, A. and {Dullemond}, C.~P. and {Kurtovic}, N.~T. and 
	{Guzm{\'a}n}, V.~V. and {Carpenter}, J.~M. and {Wilner}, D.~J. and 
	{Zhang}, S. and {Zhu}, Z. and {Birnstiel}, T. and {Bai}, X.-N. and 
	{Benisty}, M. and {Hughes}, A.~M. and {{\"O}berg}, K.~I. and 
	{Ricci}, L.},
    title = "{The Disk Substructures at High Angular Resolution Project (DSHARP). I. Motivation, Sample, Calibration, and Overview}",
  journal = {\apjl},
archivePrefix = "arXiv",
   eprint = {1812.04040},
 primaryClass = "astro-ph.SR",
     year = 2018,
    month = dec,
   volume = 869,
      eid = {L41},
    pages = {L41},
      doi = {10.3847/2041-8213/aaf741},
   adsurl = {http://adsabs.harvard.edu/abs/2018ApJ...869L..41A}
}

@ARTICLE{Vioque2018,
       author = {{Vioque}, M. and {Oudmaijer}, R.~D. and {Baines}, D. and {Mendigut{\'\i}a}, I. and {P{\'e}rez-Mart{\'\i}nez}, R.},
        title = "{Gaia DR2 study of Herbig Ae/Be stars}",
      journal = {\aap},
         year = 2018,
        month = dec,
       volume = {620},
          eid = {A128},
        pages = {A128},
          doi = {10.1051/0004-6361/201832870},
archivePrefix = {arXiv},
       eprint = {1808.00476},
 primaryClass = {astro-ph.SR},
       adsurl = {https://ui.adsabs.harvard.edu/abs/2018A&A...620A.128V}
}

@ARTICLE{Long2021,
       author = {{Long}, Feng and {Andrews}, Sean M. and {Vega}, Justin and {Wilner}, David J. and {Chandler}, Claire J. and {Ragusa}, Enrico and {Teague}, Richard and {P{\'e}rez}, Laura M. and {Calvet}, Nuria and {Carpenter}, John M. and {Henning}, Thomas and {Kwon}, Woojin and {Linz}, Hendrik and {Ricci}, Luca},
        title = "{The Architecture of the V892 Tau System: The Binary and Its Circumbinary Disk}",
      journal = {\apj},
         year = 2021,
        month = jul,
       volume = {915},
       number = {2},
          eid = {131},
        pages = {131},
          doi = {10.3847/1538-4357/abff53},
archivePrefix = {arXiv},
       eprint = {2105.02918},
 primaryClass = {astro-ph.EP},
       adsurl = {https://ui.adsabs.harvard.edu/abs/2021ApJ...915..131L}
}

@ARTICLE{Kama2016,
   author = {{Kama}, M. and {Bruderer}, S. and {Carney}, M. and {Hogerheijde}, M. and 
	{van Dishoeck}, E.~F. and {Fedele}, D. and {Baryshev}, A. and 
	{Boland}, W. and {G{\"u}sten}, R. and {Aikutalp}, A. and {Choi}, Y. and 
	{Endo}, A. and {Frieswijk}, W. and {Karska}, A. and {Klaassen}, P. and 
	{Koumpia}, E. and {Kristensen}, L. and {Leurini}, S. and {Nagy}, Z. and 
	{Perez Beaupuits}, J.-P. and {Risacher}, C. and {van der Marel}, N. and 
	{van Kempen}, T.~A. and {van Weeren}, R.~J. and {Wyrowski}, F. and 
	{Y{\i}ld{\i}z}, U.~A.},
    title = "{Observations and modelling of CO and [CI] in disks. First detections of [CI] and constraints on the carbon abundance}",
  journal = {ArXiv e-prints},
archivePrefix = "arXiv",
   eprint = {1601.01449},
 primaryClass = "astro-ph.SR",
     year = 2016,
    month = jan,
   adsurl = {http://adsabs.harvard.edu/abs/2016arXiv160101449K}
}

@ARTICLE{Hildebrand1983,
   author = {{Hildebrand}, R.~H.},
    title = "{The Determination of Cloud Masses and Dust Characteristics from Submillimetre Thermal Emission}",
  journal = {\qjras},
     year = 1983,
    month = sep,
   volume = 24,
    pages = {267},
   adsurl = {http://adsabs.harvard.edu/abs/1983QJRAS..24..267H}
}

@ARTICLE{Chiang1997,
   author = {{Chiang}, E.~I. and {Goldreich}, P.},
    title = "{Spectral Energy Distributions of T Tauri Stars with Passive Circumstellar Disks}",
  journal = {\apj},
   eprint = {astro-ph/9706042},
     year = 1997,
    month = nov,
   volume = 490,
    pages = {368-376},
   adsurl = {http://adsabs.harvard.edu/abs/1997ApJ...490..368C}
}

@ARTICLE{Bisschop2006,
   author = {{Bisschop}, S.~E. and {Fraser}, H.~J. and {{\"O}berg}, K.~I. and 
	{van Dishoeck}, E.~F. and {Schlemmer}, S.},
    title = "{Desorption rates and sticking coefficients for CO and N$_{2}$ interstellar ices}",
  journal = {\aap},
   eprint = {astro-ph/0601082},
     year = 2006,
    month = apr,
   volume = 449,
    pages = {1297-1309},
      doi = {10.1051/0004-6361:20054051},
   adsurl = {http://adsabs.harvard.edu/abs/2006A%26A...449.1297B}
}

@ARTICLE{Qi2019,
       author = {{Qi}, Chunhua and {{\"O}berg}, Karin I. and {Espaillat}, Catherine C. and
         {Robinson}, Connor E. and {Andrews}, Sean M. and {Wilner}, David J. and
         {Blake}, Geoffrey A. and {Bergin}, Edwin A. and {Cleeves}, L. Ilsedore},
        title = "{Probing CO and N$_{2}$ Snow Surfaces in Protoplanetary Disks with N$_{2}$H$^{+}$ Emission}",
      journal = {\apj},
         year = "2019",
        month = "Sep",
       volume = {882},
       number = {2},
          eid = {160},
        pages = {160},
          doi = {10.3847/1538-4357/ab35d3},
archivePrefix = {arXiv},
       eprint = {1907.10647},
 primaryClass = {astro-ph.SR},
       adsurl = {https://ui.adsabs.harvard.edu/abs/2019ApJ...882..160Q}
}

@ARTICLE{Pinilla2018,
       author = {{Pinilla}, P. and {Tazzari}, M. and {Pascucci}, I. and {Youdin}, A.~N. and
         {Garufi}, A. and {Manara}, C.~F. and {Testi}, L. and
         {van der Plas}, G. and {Barenfeld}, S.~A. and {Canovas}, H. and
         {Cox}, E.~G. and {Hendler}, N.~P. and {P{\'e}rez}, L.~M. and
         {van der Marel}, N.},
        title = "{Homogeneous Analysis of the Dust Morphology of Transition Disks Observed with ALMA: Investigating Dust Trapping and the Origin of the Cavities}",
      journal = {\apj},
         year = "2018",
        month = "May",
       volume = {859},
       number = {1},
          eid = {32},
        pages = {32},
          doi = {10.3847/1538-4357/aabf94},
archivePrefix = {arXiv},
       eprint = {1804.07301},
 primaryClass = {astro-ph.EP},
       adsurl = {https://ui.adsabs.harvard.edu/abs/2018ApJ...859...32P}
}

@ARTICLE{Francis2020,
       author = {{Francis}, Logan and {van der Marel}, Nienke},
        title = "{Dust-depleted Inner Disks in a Large Sample of Transition Disks through Long-baseline ALMA Observations}",
      journal = {\apj},
         year = 2020,
        month = apr,
       volume = {892},
       number = {2},
          eid = {111},
        pages = {111},
          doi = {10.3847/1538-4357/ab7b63},
archivePrefix = {arXiv},
       eprint = {2003.00079},
 primaryClass = {astro-ph.EP},
       adsurl = {https://ui.adsabs.harvard.edu/abs/2020ApJ...892..111F}
}

@ARTICLE{Cazzoletti2018,
   author = {{Cazzoletti}, P. and {van Dishoeck}, E.~F. and {Pinilla}, P. and 
	{Tazzari}, M. and {Facchini}, S. and {van der Marel}, N. and 
	{Benisty}, M. and {Garufi}, A. and {P{\'e}rez}, L.~M.},
    title = "{Evidence for a massive dust-trapping vortex connected to spirals. Multi-wavelength analysis of the HD 135344B protoplanetary disk}",
  journal = {\aap},
archivePrefix = "arXiv",
   eprint = {1809.04160},
 primaryClass = "astro-ph.EP",
     year = 2018,
    month = nov,
   volume = 619,
      eid = {A161},
    pages = {A161},
      doi = {10.1051/0004-6361/201834006},
   adsurl = {http://adsabs.harvard.edu/abs/2018A%26A...619A.161C}
}

@ARTICLE{Yang2023,
       author = {{Yang}, Haifeng and {Fern{\'a}ndez-L{\'o}pez}, Manuel and {Li}, Zhi-Yun and {Stephens}, Ian W. and {Looney}, Leslie W. and {Lin}, Zhe-Yu Daniel and {Harrison}, Rachel},
        title = "{Turbulent Vortex with Moderate Dust Settling Probed by Scattering-induced Polarization in the IRS 48 System}",
      journal = {\apj},
         year = 2024,
        month = mar,
       volume = {963},
       number = {2},
          eid = {134},
        pages = {134},
          doi = {10.3847/1538-4357/ad2346},
archivePrefix = {arXiv},
       eprint = {2402.12662},
 primaryClass = {astro-ph.EP},
       adsurl = {https://ui.adsabs.harvard.edu/abs/2024ApJ...963..134Y}
}

@ARTICLE{Orihara2023,
       author = {{Orihara}, Ryuta and {Momose}, Munetake and {Muto}, Takayuki and {Hashimoto}, Jun and {Liu}, Hauyu Baobab and {Tsukagoshi}, Takashi and {Kudo}, Tomoyuki and {Takahashi}, Sanemichi and {Yang}, Yi and {Hasegawa}, Yasuhiro and {Dong}, Ruobing and {Konishi}, Mihoko and {Akiyama}, Eiji},
        title = "{ALMA Band 6 high-resolution observations of the transitional disk around SY Chamaeleontis}",
      journal = {\pasj},
         year = 2023,
        month = apr,
       volume = {75},
       number = {2},
        pages = {424-445},
          doi = {10.1093/pasj/psad009},
archivePrefix = {arXiv},
       eprint = {2302.05659},
 primaryClass = {astro-ph.EP},
       adsurl = {https://ui.adsabs.harvard.edu/abs/2023PASJ...75..424O}
}

@ARTICLE{Wu2017,
       author = {{Wu}, Ya-Lin and {Sheehan}, Patrick D. and {Males}, Jared R. and {Close}, Laird M. and {Morzinski}, Katie M. and {Teske}, Johanna K. and {Haug-Baltzell}, Asher and {Merchant}, Nirav and {Lyons}, Eric},
        title = "{An ALMA and MagAO Study of the Substellar Companion GQ Lup B*}",
      journal = {\apj},
         year = 2017,
        month = feb,
       volume = {836},
       number = {2},
          eid = {223},
        pages = {223},
          doi = {10.3847/1538-4357/aa5b96},
archivePrefix = {arXiv},
       eprint = {1701.07541},
 primaryClass = {astro-ph.SR},
       adsurl = {https://ui.adsabs.harvard.edu/abs/2017ApJ...836..223W}
}

@ARTICLE{Ansdell2020,
       author = {{Ansdell}, M. and {Gaidos}, E. and {Hedges}, C. and {Tazzari}, M. and {Kraus}, A.~L. and {Wyatt}, M.~C. and {Kennedy}, G.~M. and {Williams}, J.~P. and {Mann}, A.~W. and {Angelo}, I. and {D{\^u}chene}, G. and {Mamajek}, E.~E. and {Carpenter}, J. and {Esplin}, T.~L. and {Rizzuto}, A.~C.},
        title = "{Are inner disc misalignments common? ALMA reveals an isotropic outer disc inclination distribution for young dipper stars}",
      journal = {\mnras},
         year = 2020,
        month = feb,
       volume = {492},
       number = {1},
        pages = {572-588},
          doi = {10.1093/mnras/stz3361},
archivePrefix = {arXiv},
       eprint = {1912.01610},
 primaryClass = {astro-ph.EP},
       adsurl = {https://ui.adsabs.harvard.edu/abs/2020MNRAS.492..572A}
}

@ARTICLE{Cox2017,
       author = {{Cox}, Erin G. and {Harris}, Robert J. and {Looney}, Leslie W. and {Chiang}, Hsin-Fang and {Chandler}, Claire and {Kratter}, Kaitlin and {Li}, Zhi-Yun and {Perez}, Laura and {Tobin}, John J.},
        title = "{Protoplanetary Disks in {\ensuremath{\rho}} Ophiuchus as Seen from ALMA}",
      journal = {\apj},
         year = 2017,
        month = dec,
       volume = {851},
       number = {2},
          eid = {83},
        pages = {83},
          doi = {10.3847/1538-4357/aa97e2},
archivePrefix = {arXiv},
       eprint = {1711.03974},
 primaryClass = {astro-ph.SR},
       adsurl = {https://ui.adsabs.harvard.edu/abs/2017ApJ...851...83C}
}

@ARTICLE{Garufi2022,
       author = {{Garufi}, A. and {Dominik}, C. and {Ginski}, C. and {Benisty}, M. and {van Holstein}, R.~G. and {Henning}, Th. and {Pawellek}, N. and {Pinte}, C. and {Avenhaus}, H. and {Facchini}, S. and {Galicher}, R. and {Gratton}, R. and {M{\'e}nard}, F. and {Muro-Arena}, G. and {Milli}, J. and {Stolker}, T. and {Vigan}, A. and {Villenave}, M. and {Moulin}, T. and {Origne}, A. and {Rigal}, F. and {Sauvage}, J. -F. and {Weber}, L.},
        title = "{A SPHERE survey of self-shadowed planet-forming disks}",
      journal = {\aap},
         year = 2022,
        month = feb,
       volume = {658},
          eid = {A137},
        pages = {A137},
          doi = {10.1051/0004-6361/202141692},
archivePrefix = {arXiv},
       eprint = {2111.07856},
 primaryClass = {astro-ph.GA},
       adsurl = {https://ui.adsabs.harvard.edu/abs/2022A&A...658A.137G}
}

@ARTICLE{Habart2004-hd97048,
       author = {{Habart}, E. and {Testi}, L. and {Natta}, A. and {Carbillet}, M.},
        title = "{Diamonds in HD 97048: A Closer Look}",
      journal = {\apjl},
         year = 2004,
        month = oct,
       volume = {614},
       number = {2},
        pages = {L129-L132},
          doi = {10.1086/425867},
archivePrefix = {arXiv},
       eprint = {astro-ph/0409644},
 primaryClass = {astro-ph},
       adsurl = {https://ui.adsabs.harvard.edu/abs/2004ApJ...614L.129H}
}

@ARTICLE{vanBoekel2004,
       author = {{van Boekel}, R. and {Waters}, L.~B.~F.~M. and {Dominik}, C. and {Dullemond}, C.~P. and {Tielens}, A.~G.~G.~M. and {de Koter}, A.},
        title = "{Spatially and spectrally resolved 10 {\ensuremath{\mu}}m emission in Herbig Ae/Be stars}",
      journal = {\aap},
         year = 2004,
        month = apr,
       volume = {418},
        pages = {177-184},
          doi = {10.1051/0004-6361:20034331},
       adsurl = {https://ui.adsabs.harvard.edu/abs/2004A&A...418..177V}
}

@ARTICLE{Akeson2014,
       author = {{Akeson}, R.~L. and {Jensen}, E.~L.~N.},
        title = "{Circumstellar Disks around Binary Stars in Taurus}",
      journal = {\apj},
         year = 2014,
        month = mar,
       volume = {784},
       number = {1},
          eid = {62},
        pages = {62},
          doi = {10.1088/0004-637X/784/1/62},
archivePrefix = {arXiv},
       eprint = {1402.5363},
 primaryClass = {astro-ph.SR},
       adsurl = {https://ui.adsabs.harvard.edu/abs/2014ApJ...784...62A}
}

@ARTICLE{Bruderer2012,
       author = {{Bruderer}, S. and {van Dishoeck}, E.~F. and {Doty}, S.~D. and {Herczeg}, G.~J.},
        title = "{The warm gas atmosphere of the HD 100546 disk seen by Herschel. Evidence of a gas-rich, carbon-poor atmosphere?}",
      journal = {\aap},
         year = 2012,
        month = may,
       volume = {541},
          eid = {A91},
        pages = {A91},
          doi = {10.1051/0004-6361/201118218},
archivePrefix = {arXiv},
       eprint = {1201.4860},
 primaryClass = {astro-ph.SR},
       adsurl = {https://ui.adsabs.harvard.edu/abs/2012A&A...541A..91B}
}

@ARTICLE{Wolfer2023,
       author = {{W{\"o}lfer}, L. and {Facchini}, S. and {van der Marel}, N. and {van Dishoeck}, E.~F. and {Benisty}, M. and {Bohn}, A.~J. and {Francis}, L. and {Izquierdo}, A.~F. and {Teague}, R.~D.},
        title = "{Kinematics and brightness temperatures of transition discs. A survey of gas substructures as seen with ALMA}",
      journal = {\aap},
         year = 2023,
        month = feb,
       volume = {670},
          eid = {A154},
        pages = {A154},
          doi = {10.1051/0004-6361/202243601},
archivePrefix = {arXiv},
       eprint = {2208.09494},
 primaryClass = {astro-ph.EP},
       adsurl = {https://ui.adsabs.harvard.edu/abs/2023A&A...670A.154W}
}

@ARTICLE{vanderMarel2016-isot,
       author = {{van der Marel}, N. and {van Dishoeck}, E.~F. and {Bruderer}, S. and {Andrews}, S.~M. and {Pontoppidan}, K.~M. and {Herczeg}, G.~J. and {van Kempen}, T. and {Miotello}, A.},
        title = "{Resolved gas cavities in transitional disks inferred from CO isotopologs with ALMA}",
      journal = {\aap},
         year = 2016,
        month = jan,
       volume = {585},
          eid = {A58},
        pages = {A58},
          doi = {10.1051/0004-6361/201526988},
archivePrefix = {arXiv},
       eprint = {1511.07149},
 primaryClass = {astro-ph.EP},
       adsurl = {https://ui.adsabs.harvard.edu/abs/2016A&A...585A..58V}
}

@ARTICLE{Facchini2017,
       author = {{Facchini}, S. and {Pinilla}, P. and {van Dishoeck}, E.~F. and {de Juan Ovelar}, M.},
        title = "{Inferring giant planets from ALMA millimeter continuum and line observations in (transition) disks}",
      journal = {\aap},
         year = 2018,
        month = may,
       volume = {612},
          eid = {A104},
        pages = {A104},
          doi = {10.1051/0004-6361/201731390},
archivePrefix = {arXiv},
       eprint = {1710.04418},
 primaryClass = {astro-ph.EP},
       adsurl = {https://ui.adsabs.harvard.edu/abs/2018A&A...612A.104F}
}

@ARTICLE{Ligterink2023,
       author = {{Ligterink}, N.~F.~W. and {Minissale}, M.},
        title = "{Overview of desorption parameters of volatile and complex organic molecules. A systematic dig through the experimental literature}",
      journal = {\aap},
         year = 2023,
        month = aug,
       volume = {676},
          eid = {A80},
        pages = {A80},
          doi = {10.1051/0004-6361/202346436},
archivePrefix = {arXiv},
       eprint = {2306.09071},
 primaryClass = {astro-ph.GA},
       adsurl = {https://ui.adsabs.harvard.edu/abs/2023A&A...676A..80L}
}

@ARTICLE{Dullemond2001,
       author = {{Dullemond}, C.~P. and {Dominik}, C. and {Natta}, A.},
        title = "{Passive Irradiated Circumstellar Disks with an Inner Hole}",
      journal = {\apj},
         year = 2001,
        month = oct,
       volume = {560},
       number = {2},
        pages = {957-969},
          doi = {10.1086/323057},
archivePrefix = {arXiv},
       eprint = {astro-ph/0106470},
 primaryClass = {astro-ph},
       adsurl = {https://ui.adsabs.harvard.edu/abs/2001ApJ...560..957D}
}

@ARTICLE{Minissale2022,
       author = {{Minissale}, Marco and {Aikawa}, Yuri and {Bergin}, Edwin and {Bertin}, Mathieu and {Brown}, Wendy A. and {Cazaux}, Stephanie and {Charnley}, Steven B. and {Coutens}, Audrey and {Cuppen}, Herma M. and {Guzman}, Victoria and {Linnartz}, Harold and {McCoustra}, Martin R.~S. and {Rimola}, Albert and {Schrauwen}, Johanna G.~M. and {Toubin}, Celine and {Ugliengo}, Piero and {Watanabe}, Naoki and {Wakelam}, Valentine and {Dulieu}, Francois},
        title = "{Thermal Desorption of Interstellar Ices: A Review on the Controlling Parameters and Their Implications from Snowlines to Chemical Complexity}",
      journal = {ACS Earth and Space Chemistry},
         year = 2022,
        month = mar,
       volume = {6},
       number = {3},
        pages = {597-630},
          doi = {10.1021/acsearthspacechem.1c00357},
archivePrefix = {arXiv},
       eprint = {2201.07512},
 primaryClass = {astro-ph.GA},
       adsurl = {https://ui.adsabs.harvard.edu/abs/2022ESC.....6..597M}
}

@ARTICLE{Guilloteau2025,
       author = {{Guilloteau}, S. and {Denis-Alpizar}, O. and {Dutrey}, A. and {Foucher}, C. and {Gavino}, S. and {Semenov}, D. and {Pi{\'e}tu}, V. and {Chapillon}, E. and {Testi}, L. and {Dartois}, E. and {di Folco}, E. and {Furuya}, K. and {Gorti}, U. and {Grosso}, N. and {Henning}, Th. and {Hur{\'e}}, J.~M. and {Kospal}, A. and {LePetit}, F. and {Majumdar}, L. and {Nomura}, H. and {Phuong}, N.~T. and {Ruaud}, M. and {Tang}, Y.~W. and {Wolf}, S.},
        title = "{Edge-On Disk Study (EODS): I. Thermal structure of the Flying Saucer disk}",
      journal = {\aap},
         year = 2025,
        month = aug,
       volume = {700},
          eid = {L5},
        pages = {L5},
          doi = {10.1051/0004-6361/202554853},
archivePrefix = {arXiv},
       eprint = {2507.03716},
 primaryClass = {astro-ph.EP},
       adsurl = {https://ui.adsabs.harvard.edu/abs/2025A&A...700L...5G}
}

@ARTICLE{Pinte2018,
       author = {{Pinte}, C. and {M{\'e}nard}, F. and {Duch{\^e}ne}, G. and {Hill}, T. and {Dent}, W.~R.~F. and {Woitke}, P. and {Maret}, S. and {van der Plas}, G. and {Hales}, A. and {Kamp}, I. and {Thi}, W.~F. and {de Gregorio-Monsalvo}, I. and {Rab}, C. and {Quanz}, S.~P. and {Avenhaus}, H. and {Carmona}, A. and {Casassus}, S.},
        title = "{Direct mapping of the temperature and velocity gradients in discs. Imaging the vertical CO snow line around IM Lupi}",
      journal = {\aap},
         year = 2018,
        month = jan,
       volume = {609},
          eid = {A47},
        pages = {A47},
          doi = {10.1051/0004-6361/201731377},
archivePrefix = {arXiv},
       eprint = {1710.06450},
 primaryClass = {astro-ph.SR},
       adsurl = {https://ui.adsabs.harvard.edu/abs/2018A&A...609A..47P}
}

@ARTICLE{Gavino2021,
       author = {{Gavino}, S. and {Dutrey}, A. and {Wakelam}, V. and {Guilloteau}, S. and {Kobus}, J. and {Wolf}, S. and {Iqbal}, W. and {Di Folco}, E. and {Chapillon}, E. and {Pi{\'e}tu}, V.},
        title = "{Impact of size-dependent grain temperature on gas-grain chemistry in protoplanetary disks: The case of low-mass star disks}",
      journal = {\aap},
         year = 2021,
        month = oct,
       volume = {654},
          eid = {A65},
        pages = {A65},
          doi = {10.1051/0004-6361/202038788},
archivePrefix = {arXiv},
       eprint = {2106.05888},
 primaryClass = {astro-ph.GA},
       adsurl = {https://ui.adsabs.harvard.edu/abs/2021A&A...654A..65G}
}

@ARTICLE{Yoffe2023,
       author = {{Yoffe}, G. and {van Boekel}, R. and {Li}, A. and {Waters}, L.~B.~F.~M. and {Maaskant}, K. and {Siebenmorgen}, R. and {van den Ancker}, M. and {Petit dit de la Roche}, D.~J.~M. and {Lopez}, B. and {Matter}, A. and {Varga}, J. and {Hogerheijde}, M.~R. and {Weigelt}, G. and {Oudmaijer}, R.~D. and {Pantin}, E. and {Meyer}, M.~R. and {Augereau}, J. -C. and {Henning}, Th.},
        title = "{Spatially resolving polycyclic aromatic hydrocarbons in Herbig Ae disks with VISIR-NEAR at the VLT}",
      journal = {\aap},
         year = 2023,
        month = jun,
       volume = {674},
          eid = {A57},
        pages = {A57},
          doi = {10.1051/0004-6361/202245656},
archivePrefix = {arXiv},
       eprint = {2303.06592},
 primaryClass = {astro-ph.EP},
       adsurl = {https://ui.adsabs.harvard.edu/abs/2023A&A...674A..57Y}
}

@INPROCEEDINGS{Alexander2008,
       author = {{Alexander}, Conel M. O'D. and {Cody}, George D. and {Fogel}, Marilyn and {Yabuta}, Hikaru},
        title = "{Organics in meteorites - Solar or interstellar?}",
    booktitle = {Organic Matter in Space},
         year = 2008,
       editor = {{Kwok}, Sun and {Sanford}, Scott},
       series = {IAU Symposium},
       volume = {251},
        month = oct,
        pages = {293-298},
          doi = {10.1017/S1743921308021765},
       adsurl = {https://ui.adsabs.harvard.edu/abs/2008IAUS..251..293A}
}

@ARTICLE{Kelly2007,
       author = {{Kelly}, Brandon C.},
        title = "{Some Aspects of Measurement Error in Linear Regression of Astronomical Data}",
      journal = {\apj},
         year = 2007,
        month = aug,
       volume = {665},
       number = {2},
        pages = {1489-1506},
          doi = {10.1086/519947},
archivePrefix = {arXiv},
       eprint = {0705.2774},
 primaryClass = {astro-ph},
       adsurl = {https://ui.adsabs.harvard.edu/abs/2007ApJ...665.1489K}
}

@ARTICLE{Sturm2024,
       author = {{Sturm}, J.~A. and {McClure}, M.~K. and {Harsono}, D. and {Bergner}, J.~B. and {Dartois}, E. and {Boogert}, A.~C.~A. and {Cordiner}, M.~A. and {Drozdovskaya}, M.~N. and {Ioppolo}, S. and {Law}, C.~J. and {Lis}, D.~C. and {McGuire}, B.~A. and {Melnick}, G.~J. and {Noble}, J.~A. and {{\"O}berg}, K.~I. and {Palumbo}, M.~E. and {Pendleton}, Y.~J. and {Perotti}, G. and {Rocha}, W.~R.~M. and {Urso}, R.~G. and {van Dishoeck}, E.~F.},
        title = "{A JWST/MIRI analysis of the ice distribution and polycyclic aromatic hydrocarbon emission in the protoplanetary disk HH 48 NE}",
      journal = {\aap},
         year = 2024,
        month = sep,
       volume = {689},
          eid = {A92},
        pages = {A92},
          doi = {10.1051/0004-6361/202450865},
archivePrefix = {arXiv},
       eprint = {2407.09627},
 primaryClass = {astro-ph.EP},
       adsurl = {https://ui.adsabs.harvard.edu/abs/2024A&A...689A..92S}
}

@ARTICLE{Tripathi2017,
       author = {{Tripathi}, Anjali and {Andrews}, Sean M. and {Birnstiel}, Tilman and {Wilner}, David J.},
        title = "{A millimeter Continuum Size-Luminosity Relationship for Protoplanetary Disks}",
      journal = {\apj},
         year = 2017,
        month = aug,
       volume = {845},
       number = {1},
          eid = {44},
        pages = {44},
          doi = {10.3847/1538-4357/aa7c62},
archivePrefix = {arXiv},
       eprint = {1706.08977},
 primaryClass = {astro-ph.EP},
       adsurl = {https://ui.adsabs.harvard.edu/abs/2017ApJ...845...44T}
}

@ARTICLE{Pinilla2015,
       author = {{Pinilla}, P. and {Birnstiel}, T. and {Walsh}, C.},
        title = "{Sequential planet formation in the HD 100546 protoplanetary disk?}",
      journal = {\aap},
         year = 2015,
        month = aug,
       volume = {580},
          eid = {A105},
        pages = {A105},
          doi = {10.1051/0004-6361/201425539},
archivePrefix = {arXiv},
       eprint = {1506.02383},
 primaryClass = {astro-ph.EP},
       adsurl = {https://ui.adsabs.harvard.edu/abs/2015A&A...580A.105P}
}

@ARTICLE{vanderMarel2019,
       author = {{van der Marel}, Nienke and {Dong}, Ruobing and {di Francesco}, James and {Williams}, Jonathan P. and {Tobin}, John},
        title = "{Protoplanetary Disk Rings and Gaps across Ages and Luminosities}",
      journal = {\apj},
         year = 2019,
        month = feb,
       volume = {872},
       number = {1},
          eid = {112},
        pages = {112},
          doi = {10.3847/1538-4357/aafd31},
archivePrefix = {arXiv},
       eprint = {1901.03680},
 primaryClass = {astro-ph.EP},
       adsurl = {https://ui.adsabs.harvard.edu/abs/2019ApJ...872..112V}
}

@ARTICLE{vanderMarel2018,
       author = {{van der Marel}, Nienke and {Williams}, Jonathan P. and {Ansdell}, M. and {Manara}, Carlo F. and {Miotello}, Anna and {Tazzari}, Marco and {Testi}, Leonardo and {Hogerheijde}, Michiel and {Bruderer}, Simon and {van Terwisga}, Sierk E. and {van Dishoeck}, Ewine F.},
        title = "{New Insights into the Nature of Transition Disks from a Complete Disk Survey of the Lupus Star-forming Region}",
      journal = {\apj},
         year = 2018,
        month = feb,
       volume = {854},
       number = {2},
          eid = {177},
        pages = {177},
          doi = {10.3847/1538-4357/aaaa6b},
archivePrefix = {arXiv},
       eprint = {1801.06154},
 primaryClass = {astro-ph.EP},
       adsurl = {https://ui.adsabs.harvard.edu/abs/2018ApJ...854..177V}
}

@ARTICLE{Lagage2006,
       author = {{Lagage}, Pierre-Olivier and {Doucet}, Coralie and {Pantin}, Eric and {Habart}, Emilie and {Duch{\^e}ne}, Gaspard and {M{\'e}nard}, Fran{\c{c}}ois and {Pinte}, Christophe and {Charnoz}, S{\'e}bastien and {Pel}, Jan-Willem},
        title = "{Anatomy of a Flaring Proto-Planetary Disk Around a Young Intermediate-Mass Star}",
      journal = {Science},
         year = 2006,
        month = oct,
       volume = {314},
       number = {5799},
        pages = {621-623},
          doi = {10.1126/science.1131436},
       adsurl = {https://ui.adsabs.harvard.edu/abs/2006Sci...314..621L}
}

@ARTICLE{Maaskant2014,
       author = {{Maaskant}, K.~M. and {Min}, M. and {Waters}, L.~B.~F.~M. and {Tielens}, A.~G.~G.~M.},
        title = "{Polycyclic aromatic hydrocarbon ionization as a tracer of gas flows through protoplanetary disk gaps}",
      journal = {\aap},
         year = 2014,
        month = mar,
       volume = {563},
          eid = {A78},
        pages = {A78},
          doi = {10.1051/0004-6361/201323137},
archivePrefix = {arXiv},
       eprint = {1402.0902},
 primaryClass = {astro-ph.SR},
       adsurl = {https://ui.adsabs.harvard.edu/abs/2014A&A...563A..78M}
}

@ARTICLE{Booth2025,
       author = {{Booth}, Alice S. and {W{\"o}lfer}, Lisa and {Temmink}, Milou and {Calahan}, Jenny and {Evans}, Lucy and {Law}, Charles J. and {Leemker}, Margot and {Notsu}, Shota and {{\"O}berg}, Karin and {Walsh}, Catherine},
        title = "{Ice Sublimation in the Dynamic HD 100453 Disk Reveals a Rich Reservoir of Inherited Complex Organics}",
      journal = {\apjl},
         year = 2025,
        month = jun,
       volume = {986},
       number = {1},
          eid = {L9},
        pages = {L9},
          doi = {10.3847/2041-8213/adc7b2},
archivePrefix = {arXiv},
       eprint = {2504.14023},
 primaryClass = {astro-ph.EP},
       adsurl = {https://ui.adsabs.harvard.edu/abs/2025ApJ...986L...9B}
}

@ARTICLE{Bouteraon2019,
       author = {{Bout{\'e}raon}, T. and {Habart}, E. and {Ysard}, N. and {Jones}, A.~P. and {Dartois}, E. and {Pino}, T.},
        title = "{Carbonaceous nano-dust emission in proto-planetary discs: the aliphatic-aromatic components}",
      journal = {\aap},
         year = 2019,
        month = mar,
       volume = {623},
          eid = {A135},
        pages = {A135},
          doi = {10.1051/0004-6361/201834016},
archivePrefix = {arXiv},
       eprint = {1901.07332},
 primaryClass = {astro-ph.GA},
       adsurl = {https://ui.adsabs.harvard.edu/abs/2019A&A...623A.135B}
}

@ARTICLE{vanderMarel2021-asymm,
       author = {{van der Marel}, Nienke and {Birnstiel}, Til and {Garufi}, Antonio and {Ragusa}, Enrico and {Christiaens}, Valentin and {Price}, Daniel J. and {Sallum}, Steph and {Muley}, Dhruv and {Francis}, Logan and {Dong}, Ruobing},
        title = "{On the Diversity of Asymmetries in Gapped Protoplanetary Disks}",
      journal = {\aj},
         year = 2021,
        month = jan,
       volume = {161},
       number = {1},
          eid = {33},
        pages = {33},
          doi = {10.3847/1538-3881/abc3ba},
archivePrefix = {arXiv},
       eprint = {2010.10568},
 primaryClass = {astro-ph.EP},
       adsurl = {https://ui.adsabs.harvard.edu/abs/2021AJ....161...33V}
}

@ARTICLE{Zhu2019,
       author = {{Zhu}, Zhaohuan and {Zhang}, Shangjia and {Jiang}, Yan-Fei and {Kataoka}, Akimasa and {Birnstiel}, Tilman and {Dullemond}, Cornelis P. and {Andrews}, Sean M. and {Huang}, Jane and {P{\'e}rez}, Laura M. and {Carpenter}, John M. and {Bai}, Xue-Ning and {Wilner}, David J. and {Ricci}, Luca},
        title = "{One Solution to the Mass Budget Problem for Planet Formation: Optically Thick Disks with Dust Scattering}",
      journal = {\apjl},
         year = 2019,
        month = jun,
       volume = {877},
       number = {2},
          eid = {L18},
        pages = {L18},
          doi = {10.3847/2041-8213/ab1f8c},
archivePrefix = {arXiv},
       eprint = {1904.02127},
 primaryClass = {astro-ph.EP},
       adsurl = {https://ui.adsabs.harvard.edu/abs/2019ApJ...877L..18Z}
}

@ARTICLE{Leemker2022,
       author = {{Leemker}, M. and {Booth}, A.~S. and {van Dishoeck}, E.~F. and {P{\'e}rez-S{\'a}nchez}, A.~F. and {Szul{\'a}gyi}, J. and {Bosman}, A.~D. and {Bruderer}, S. and {Facchini}, S. and {Hogerheijde}, M.~R. and {Paneque-Carre{\~n}o}, T. and {Sturm}, J.~A.},
        title = "{Gas temperature structure across transition disk cavities}",
      journal = {\aap},
         year = 2022,
        month = jul,
       volume = {663},
          eid = {A23},
        pages = {A23},
          doi = {10.1051/0004-6361/202243229},
archivePrefix = {arXiv},
       eprint = {2204.03666},
 primaryClass = {astro-ph.EP},
       adsurl = {https://ui.adsabs.harvard.edu/abs/2022A&A...663A..23L}
}

\appendix
\onecolumn 
\section{ALMA gallery}
This Appendix contains the ALMA images analyzed in this work.

\begin{figure*}
    \includegraphics[width=\textwidth]{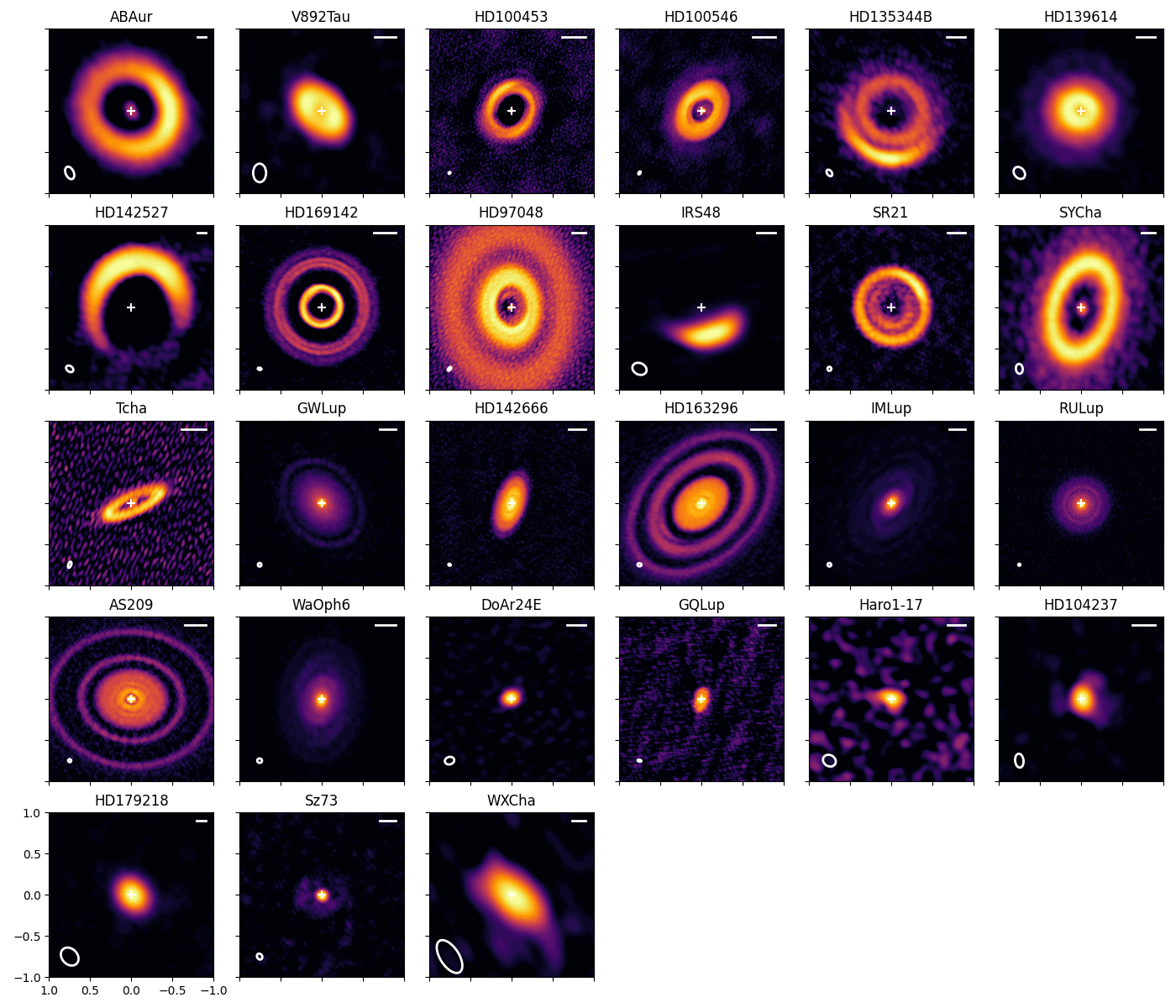}
\caption{Gallery ALMA continuum images of the targets in this study. The images are shown with a arcsinh stretch to enhance faint features. The stellar position is indicated with a plus sign and the beam size is shown in the lower left corner of each image. The images are 2"$\times$2" except for HD142527, which is 4"$\times$4" due to the larger ring size.}
\label{fig:gallery}
\end{figure*}

\clearpage
\newpage
\section{SEDs of transition disks}
\label{sec:SEDs}
In this section we present the Spectral Energy Distributions (SEDs) and mid-plane temperature profiles of the transition disks in this study based on RADMC-3D modeling, since their temperature cannot be estimated with a simple power-law. 

For HD135344B, HD169142 and HD97048, SED radiative transfer modeling was performed by \citet{vanderMarel2019}, and we use their derived temperature profiles. For IRS48, we use the mid-plane temperature profile from \citet{vanderMarel2021-irs48} and for HD142527, we use the profile from \citet{Temmink2023}.

For the other transition disks, no SED fits with temperature profiles were available in the literature, so we fit RADMC-3D models ourselves. The SEDs were constructed using photometry from Johnson BVR, Gaia, 2MASS, \emph{WISE}, \emph{Spitzer}, \emph{IRAS}, \emph{Herschel/PACS} and ALMA millimeter fluxes from this study. The optical-near infrared photometric data points were corrected for extinction using the $A_V$ values from \citet{Francis2020}. For the fitting itself we follow a similar procedure as \citet[][Appendix E]{Temmink2023}, but with a power-law surface density with an empty cavity inside cavity radius $r_{\rm cav}$ and an inner dust disk of 1 au which is depleted by a factor $\delta_{\rm dust}$. The outer radius is set at 150 au and the surface density is scaled to the total gas disk mass $M_{\rm disk}$ with a gas-to-dust ratio of 100. The dust opacities are computed using a combination of a small grain and large grain population, of 0.005-1 $\mu$m and 0.005-1000 $\mu$m, respectively. The vertical scale height is described by $h(r)=h_c({r}/{r_c})^{\psi}$ with $\psi=0.15$ and characteristic radius $r_c$=50 au. Stellar parameters are taken from Table \ref{tbl:sample}.

The free parameters in the model are thus the total disk mass $M_{\rm disk}$, the vertical height $h_c$ at characteristic radius $r_c$, the cavity radius $r_{\rm cav}$ and inner disk depletion factor $\delta_{\rm dust}$. The models were fit to both the radial continuum profiles and the SEDs. The final SEDs with their best fit models are shown in Figure \ref{fig:seds} and the best-fit parameters are given in Table \ref{tbl:sedfits}. For T Cha, the SED fit had to be run at lower inclination than for the continuum image due to the obscuration of the star by the inner disk, indicating that the inner disk may be misaligned with the outer disk, similar to RY Lup \citep{vanderMarel2018}.

\begin{table}[!ht]
    \caption{Best-fit parameters SEDs}
    \label{tbl:sedfits}
    \centering
    \begin{tabular}{l|llll}
    \hline
         Target & $M_{\rm disk}^a$ & $h_c$ & $r_{\rm cav}$ & $\delta_{\rm dust}$\\ 
         &($10^{-3} M_{\odot}$)&&(au)& \\
    \hline     
         V892Tau & 4 &0.05&23&10$^{-3}$ \\ 
         HD100453 & 2&0.20&32&10$^{-2}$  \\ 
         HD100546 & 8&0.20&30&10$^{-1}$  \\ 
         HD139614 & 6.6 &0.10&13&10$^{0}$  \\ 
         SR21 & 4&0.15&60&10$^{-3}$  \\ 
         SYCha & 10 &0.10&106&10$^{0}$  \\ 
         TCha & 10 &0.10&28&10$^{-1}$  \\ 
    \hline     
    \end{tabular}\\
    $^a$) $M_{\rm disk}$ is the total disk gas mass: the disk dust mass is computed within RADMC-3D assuming a gas-to-dust ratio of 100.
\end{table}

\begin{figure*}[!h]
    \includegraphics[width=0.9\textwidth]{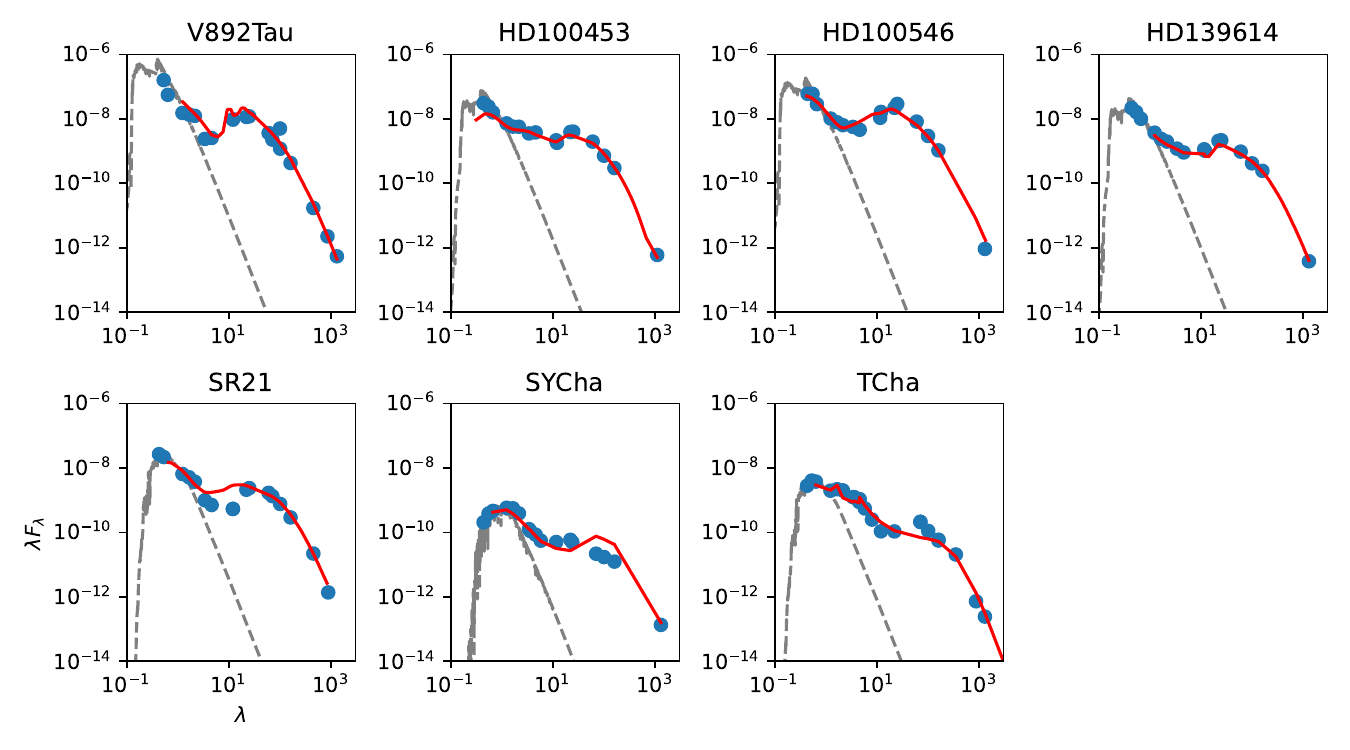}
\caption{SEDs of the transition disks for which radiative transfer models were run. The plots show the best-fit models and photometry.}
\label{fig:seds}
\end{figure*}

\begin{figure*}[!h]
    \includegraphics[width=0.9\textwidth]{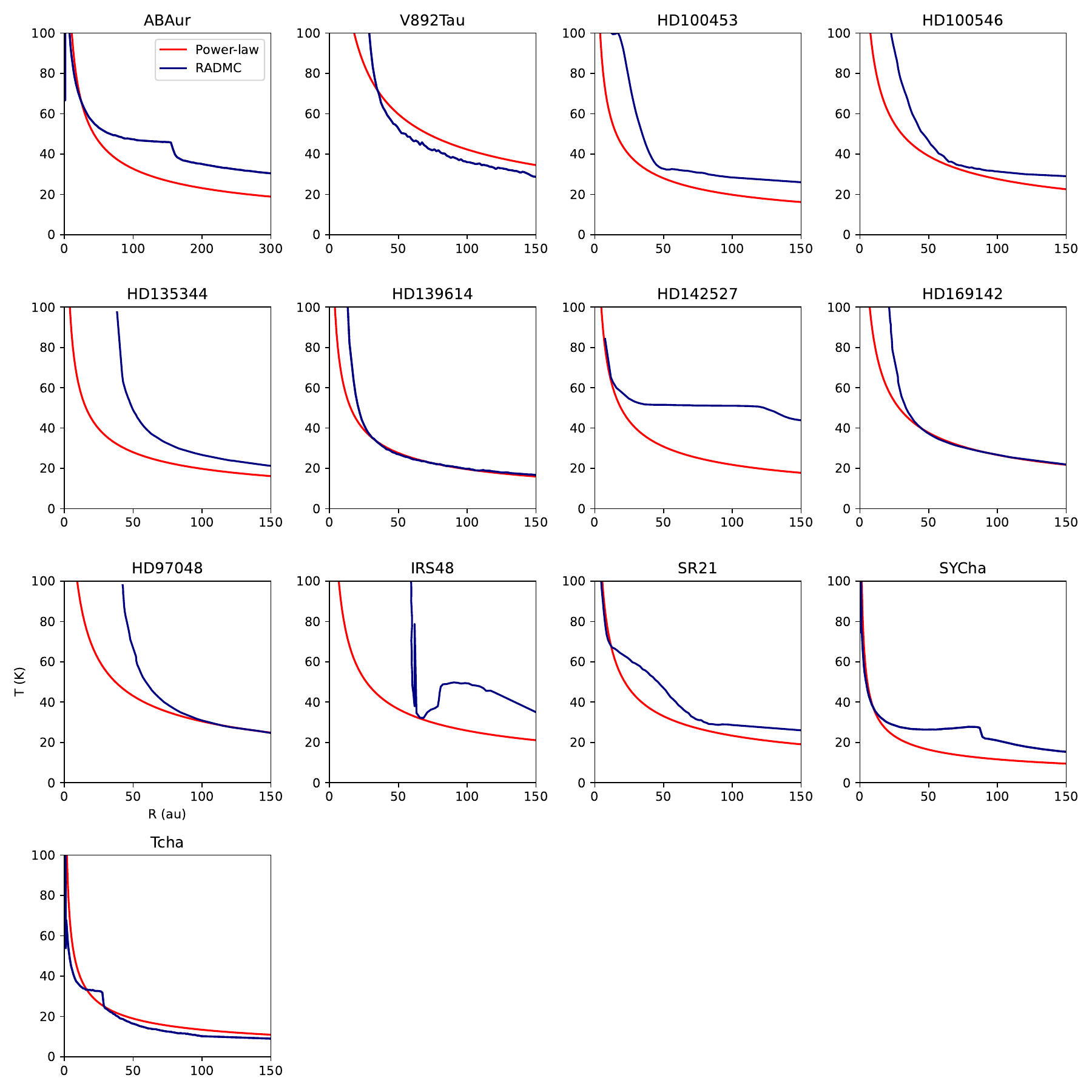}
\caption{Temperature profiles of the best-fit radiative transfer models to the SEDs of the transition disks in this study. The red curve in each plot shows the expected temperature profile according to the power-law estimate. This is generally an underestimate of the actual temperature due to the irradiated cavity wall.}
\label{fig:temperaturetds}
\end{figure*}

\end{document}